\documentclass[journal=jacsat,manuscript=article]{achemso}

\usepackage[version=3]{mhchem} 
\usepackage[english]{babel}
\usepackage[utf8]{inputenc}
\usepackage{graphicx}
\usepackage{SIunits}
\usepackage[nottoc]{tocbibind}
\usepackage{natbib}
\usepackage{xcolor}
\usepackage{caption}
\usepackage{tabularx}
\DeclareCaptionLabelFormat{adja-page}{\hrulefill\\#1 #2 \emph{(previous page)}}

\author{Firoozeh Babayekhorasani}
\affiliation[UNSW]
{School of Chemical Engineering, University of New South Wales, Sydney, NSW, Australia}

\author{Maryam Hosseini}
\affiliation[UNSW]
{School of Chemical Engineering, University of New South Wales, Sydney, NSW, Australia}

\author{Patrick T. Spicer}
\affiliation[UNSW]
{School of Chemical Engineering, University of New South Wales, Sydney, NSW, Australia}

\email{p.spicer@unsw.edu.au }
\phone{+61 2 9385 5744}

\title[An \textsf{achemso} demo]
  {Molecular and colloidal transport in bacterial cellulose hydrogels}


\begin{document}

\begin{tocentry}
\center
\includegraphics[scale=1]{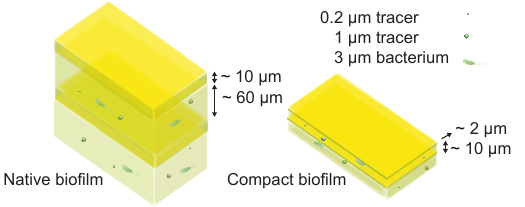}
\end{tocentry}

\begin{abstract}
 Bacterial cellulose biofilms are 
 complex  networks 
 of strong interwoven nanofibers that control 
 transport and protect bacterial 
 colonies in the film. 
 Design of diverse 
 applications of these bacterial cellulose 
 films also relies on understanding 
 and controlling transport through 
 the fiber mesh, 
 and transport simulations of the films 
 are most accurate when guided 
 by experimental characterization of the 
 structures and the 
 resultant diffusion inside.
 Diffusion through such films is a 
 function of their key microstructural
 length scales, determining how molecules, 
 as well as particles and microorganisms, 
 permeate them. 
 We  use  microscopy  
 to  study the unique bacterial cellulose 
 film via its pore structure and 
 quantify the  mobility 
 dynamics  of various sizes of tracer 
 particles and macromolecules. 
 Mobility is hindered within the films, 
 as confinement and local movement strongly 
 depend on the void size  
 relative to diffusing tracers.  
 The biofilms have a naturally periodic structure of 
 alternating dense and porous 
 layers of nanofiber mesh, 
 and we tune the magnitude of the spacing via 
 fermentation conditions.
 Micron-sized particles can diffuse
 through the porous layers, 
 but can not penetrate the dense layers. 
Tracer mobility in the porous layers 
 is isotropic, indicating a largely 
 random pore structure there. 
 Molecular diffusion through the whole film 
 is only slightly reduced  
 by the structural tortuosity. 
 Knowledge of transport variations 
 within bacterial cellulose networks 
 can be used to guide the design of symbiotic 
 cultures in these structures and enhance 
 their use in applications like biomedical 
 implants, wound dressings, lab-grown meat, 
 clothing textiles, and sensors.
 
\end{abstract}


\section{keywords}
cellulose hydrogels, diffusion, microscopy, confined mobility, biofilms
\section{Introduction}
Bacterial cellulose is produced in a 
film at the air-liquid interface of 
\textit{Acetobacter xylinum}  
cultures that transform 
aqueous sugar into balsamic vinegar 
and kombucha 
\cite{soares2021technological,chakravorty2016}, 
but the cellulose itself is an 
active area of exploration by researchers. 
In contrast with biofilms based 
on cell-produced exopolysaccharides 
(EPS), here bacteria build 
a network of pure cellulose 
fibers of nanometer-scale thickness 
and micron-scale lengths 
intertwined in a mesh 
structure, or pellicle: 
a stable living 
environment for symbiotic 
cultures of bacteria and 
yeast (SCOBY).  \cite{rooney2020} 
Bacterial cellulose biofilms are remarkably 
strong \cite{Tanpichai2012} versus 
the softer biofilms made of 
EPS, prompting their study for use 
as packaging \cite{shi2014}, 
biomedical implants 
\cite{bangasser2017, witzler2018, yamada2019}, 
wound dressings \cite{markstedt2015, curvello2019, Ahmed2020}, 
sensors \cite{gilbert2021living}, 
batteries \cite{Chen2018a,Kim2019}, 
food \cite{schaffner2017,hausmann2018,joshi2018},
clothing \cite{quijano2021future,da2021bacterial}
and 
lab-grown meat \cite{rybchyn2021nanocellulose}.
All of these applications benefit 
from a complex cellulose fiber mesh 
providing mechanical strength, low 
density, and high surface area, 
but their performance is 
determined by the transport 
properties inside the mesh 
\cite{Witten2017}. Modifying 
the pore size of the film or surface 
properties of the cellulose nanofibers, 
for example, could enhance cell 
retention and adherence for 
lab-grown meat 
production \cite{rybchyn2021nanocellulose}. 
In order to accurately design 
and simulate\cite{stylianopoulos2008permeability} materials made from 
bacterial cellulose films, a 
careful characterization of their 
structure and transport properties 
is needed, and that is the focus 
of this work.

In random fiber networks, mesh size, 
alignment and heterogeneity can lead to 
strong variations in microparticle transport rates \cite{Stylianopoulos2010,sykes2016tailoring,Chauhan2011} 
and localization \cite{Sentjabrskaja2016,Poling-Skutvik2019}. 
Particle dynamics are coupled to any 
spatial heterogeneity of the matrix \cite{roberts2019tracer}, 
for example, affecting drug transport in cancer 
tissue\cite{Stylianopoulos2010,Jain2010} 
as well as the movement of soft particle medical delivery 
vehicles \cite{yu2018rapid,Yu2019,Bao2020}. 
Complex biological network structures can vary 
significantly\cite{Huck2019}, but  
diffusive transport can be measured in the matrix via 
molecular or colloidal tracer studies \cite{birjiniuk2014single, billings2015material, Mastorakos2015, Marczynski2018, dunsing2019purely}. 
Diffusive transport in EPS biofilm networks 
has been studied to measure how 
rheology \cite{stewart2015artificial,ganesan2016associative},  
diffusive permeability \cite{kundukad2017mechanistic,boudarel2021situ}, 
and restructuring modulate transport  
in the matrix \cite{chew2014dynamic, kundukad2016mechanical}, 
but cellulose biofilms have not been characterized in the same way. 
In this study, we use optical microscopy to track 
molecular and particulate \cite{Crocker1996} tracers  
to characterize mobility 
within anisotropic, hierarchically structured 
bacterial cellulose films 
and show the diversity of 
natural structures and properties 
found there.  

We find bacterial cellulose 
films naturally 
exhibit strong periodic 
variations in fiber packing 
density over micron-scale distances, alternating 
between porous layers that are permeable to cells or colloids
and dense layers that are impermeable to all but molecular transport.
We tune the overall density of the films 
by adjusting the viscosity of the growth medium, demonstrating its effects on diffusive transport. 
Diffusion of colloidal tracer particles in the films depends on their size 
relative to the fiber mesh dimensions and heterogeneity. 
Mean squared displacement of a range of particle sizes is sub-diffusive 
and non-Gaussian in the films, and the transition from diffusive to 
anomalous sub-diffusive behavior is determined by the size of the 
particle relative to the 
mesh void space and the time scale of 
interaction. 
We also observe that the 
diffusive dynamics of macromolecules are 
mostly isotropic in the films, 
despite the random fiber structure. 
Understanding transport within 
bacterial cellulose films 
is necessary to 
enable rational design and simulation 
of myriad biomedical and synthetic materials, 
as they all depend on transport of molecules, 
particles, or microorganisms through 
their structures for viability and performance.

\section{Experimental}

\subsection*{Preparation of the bacterial biofilms}

\textit{Acetobacter xylinum} was provided by Nourishme Organics, Australia, and used to grow the pellicles. The culture medium was prepared by seeping black tea into boiling water for 10 minutes. Sucrose solutions at 10 \% w/v were added to the tea solution. 

The culture medium was then mixed with aqueous sodium alginate solution of different concentrations to make the final concentration of the cultures between 0 - 1 \% w/v. 
The culture medium was incubated at $25$ $^{\circ}$C for $10$ days. Cellulose fibers formed in the culture and moved to the air-liquid interface. As more fibers forms, the thickness of the layer at the interface increases. After 10 days, the consolidated films were collected from the interface. The film was carefully washed using DI water to remove bacteria and any loosely attached fibrous material from the films. The films were washed in $0.5$ $M$ NaOH (Chem-Supply Pty Ltd, Australia) at $90$ $^{\circ}$C for $90$ minutes to kill the remaining bacteria. The films were then washed with water until neutralized. The washed films were immersed in an aqueous solution of $1.5$ $\%$ $w/v$ methylchloroisothiazolinone preservative (Sigma Aldrich, Sydney, Australia) and stored at $4$ $^{\circ}$C before the microscopy experiments.

\subsection*{Characterization of the bacterial biofilms}

The structure of the biofilms is characterized using scanning electron microscopy (SEM), confocal, and light-sheet microscopy techniques. Samples are freeze-dried overnight and imaged from the side interface using a field-emission scanning electron microscope (NanoSEM 230). 

The side interface of the wet samples is also imaged using a ZEISS Light-sheet $Z.1$ microscope with a $20\times$ water immersion objective (NA=1.0). To image the samples using the light-sheet microscope, the biofilms are first stained using $0.5$ \% $w/v$ Congo Red dye (Sigma Aldrich, Sydney, Australia) and incubated for $15$ minutes. The samples are cut into small rectangular cubes with dimensions of roughly $1$ $mm$ $\times$ $5$ $mm$ $\times$ the film height. The samples are placed inside PTFE tubing with a matched refractive index to water. The tube is filled with deionized (DI) water. Both ends of the tube are sealed with PARAFILM (Sigma Aldrich, Sydney, Australia). The tube is then connected to a syringe and is loaded inside the microscope chamber. A thin section of the sample is imaged by illuminating from the left and right sides. The two sides are then combined using an automated dual side fusion method by ZEN Black software.

\subsection*{Preparation of tracer particle dispersions in the biofilms}
Polybead carboxylate green dyed microspheres with diameter of 0.20 $\mu m$, 0.50 $\mu m$, and 1 $\mu m$ at a concentration of 2.5 \% were purchased from Polyscience, Inc (Warrington, PA, USA) and were diluted using DI water, 1000$\times$, 500$\times$, and 200$\times$, respectively. The homogenized particle suspension is added to the films that are already stained with Congo red dye ($0.5$ \% $w/v$) and incubated overnight to allow particle diffusion through the biofilm matrix. 

\subsection*{Imaging of the tracer particles diffusing through the biofilms}

The samples were imaged using a light-sheet microscope equipped with a 20$\times$ water immersion objective of numerical aperture NA=1.0 with a frame rate of 5 fps from the side interface. To ensure the particles are diffusing through the biofilm, we imaged both the biofilm matrix and the particles simultaneously using two separate channels. We acquired time-series images of different locations through the depth of the side interface of the biofilms. We acquired 1000 images over time with pixel size of 3.94 pix/$\mu m$ and image size of 487 $\times$ 487 $\mu m^2$. We used single-particle tracking algorithm \cite{Crocker1996} to locate and track the particles over time. Using particle trajectories, we calculated the ensemble average mean squared displacement, \( \Delta x^2 (\Delta t) = \big <(x(t+\Delta t) -x(t))^2 \big>\), over lag time, $\Delta t$.

\subsection*{Fluorescence Recovery After Photo-bleaching (FRAP) analysis}
Recovery of the intensity of fluorescein isothiocyanate–dextran (FITC-Dextran) with three molecular weights of 4, 20, 40 kDa was measured using Zeiss LSM 880 (JENA/Germany) confocal microscopes. The film is first cut into a regular shape with 
dimensions of 5 $\times$ 5 $\times$ 2  $mm^3$ from the side interface of the full film. 
The Congo Red labeled film was incubated in 1 mg/ml of FITC-Dextran solutions 
for 24 hr under gentle agitation. The film is imaged using a 40X water immersion objective (NA=1.1). The selected area is then bleached for about 2$-$3 sec and imaged over 1000 frames 
to follow the recovery of intensity with a frame rate of 30 fps.

\section{Results and discussion}
\begin{figure}
\center
\includegraphics[scale=1.15]{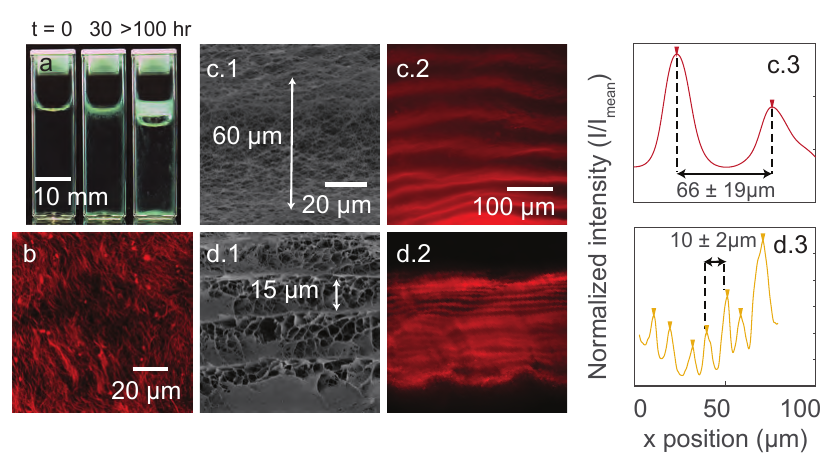}

\caption{Macro and microstructure of the cellulose pellicle film. a) A cellulose biofilm forms as a white layer at the top of a cuvette during bacterial production of nanofibers. b) Microstructure of the film imaged from the top interface. c) SEM images (c.1), light-sheet micrograph (c.2) and intensity profile measured from the light-sheet image (c.3) of the side of the film grown in the media with no sodium alginate. d) SEM images (d.1), light-sheet micrograph (d.2) and intensity profile measured from the light-sheet image (d.3) of the side of the film grown in the media with $1$ \% $w/v$ of sodium alginate. The unique periodic spacing of denser layers within the overall fiber mesh structure is visible from the side interface.}%
\label{fig:biofilms}%
\end{figure}

Cellulose biofilms are formed during static growth of \emph{Acetobacter xylinum} in an 
aqueous culture, as the cellulose fibers gradually combine and 
move up to the liquid-air interface. 
Over time, more fibers grow and join the layer on the top and produce a 
consolidated pellicle film (Figure \ref{fig:biofilms}a). 
The film is an entangled network of cellulose fibers, seen when the 
structure is imaged from its top or bottom interface (Figure \ref{fig:biofilms}b). 
The network shows a distinct degree of structural periodicity when observed from its 
side (Figure \ref{fig:biofilms} c.1$-$c.3) where dense layers of fibers  
alternate with more porous regions of 
random fibrous mesh (Figure \ref{fig:biofilms} c.1$-$c.3). 
The regions are all sufficiently porous to allow 
nutrient transfer through the biofilm, but the denser intermediate 
layers are likely to restrict the diffusion of most particles or cells \cite{birjiniuk2014single}. 
The periodic fiber density gradient is fairly typical of cellulose 
films\cite{Bodin2007,Hu2010,Janpetch2016,gromovykh2020structural}, 
and is likely a result of periods of surface drying and fiber 
consolidation as well as different bacterial growth patterns \cite{qin2020cell}.

The periodic variation in film permeability and transport creates 
a complex pathway for diffusing species and likely aids in 
protection of the bacteria, but we are not aware of specific evolutionary benefits.  
The overall fiber density of the mesh can also be controlled, if needed, 
in applications requiring different microstructures and pore 
accessibility \cite{Fu2012,Huang2013}. One method of control is to increase
the viscosity of the culture medium by adding thickeners 
like sodium alginate,\cite{Zhou2007,Cheng2009,Kwandou2018} 
that reduce bacterial mobility throughout the culture medium. As the medium viscosity increases in the culture, the thickness of the film and the spacing between the periodic layers decreases (Figure \ref{fig:biofilms} d.1$-$d.3). Interestingly, the mass of cellulose produced does not change significantly. This indicates the growth rate of cellulose is unchanged, but the mobility of the bacteria is expected to be decreased by the increased viscosity. As a result, the bacteria do not spread out as widely but still grow the same amount of cellulose, resulting in thinner and denser films. To some extent the spacing between the periodic layers and the overall porosity of the film can be controlled by tuning the viscosity and movement of the bacteria in the culture.
The stark difference wrought by addition of the thickener is easily 
demonstrated by comparing the 
films grown without sodium alginate, Figure \ref{fig:biofilms}c, 
and with $1 \% \; w/v$ sodium alginate, Figure \ref{fig:biofilms}d.
The native structure in Figure \ref{fig:biofilms}c has 
periodic dense layers with regular spacing of around \unit{60}{\micro\meter} 
while the structure in Figure \ref{fig:biofilms}d has a layer 
spacing nearly 6$\times$ smaller. 
The two biofilms contain the same mass of cellulose, 
but their varied structures result in respective bulk fiber 
densities of only \unit{0.01}{\gram\per\centi\cubic\meter} 
and \unit{0.03}{\gram\per\centi\cubic\meter} based on dry 
weight and wet volume measurements.
The results in Figure  \ref{fig:biofilms} demonstrate that 
we can control the physical spacing and relative 
distance within these periodic fiber structures 
to understand and control diffusive transport effects.
Below we study the effects of the varied spacing and pore size 
distribution on directional transport using particle tracking.

We study the mobility 
of tracers that range from macromolecular 
to colloidal length scales, \unit{2}{\nano\meter} $-$ \unit{1}{\micro\meter}, 
because the two different films have pores that span similar dimensions. 
Combined light-sheet and confocal microscopy techniques 
enable structural quantification of large biological samples \cite{sankaran2019single,qin2020cell}
and single particle tracking \cite{Crocker1996}. 
Fluorescence Recovery After Photobleaching (FRAP) resolves 
molecular-scale transport within the overall network. 
As the cellulose fibers are strongly interwoven in the mesh due to  hydrogen bonding, they won’t be able to move or fluctuate noticeably. Imaging the stained fibers simultaneously with particles over time did not show any movement by fibers over the course of the experiments.

\begin{figure}
\center
\includegraphics[scale=0.85]{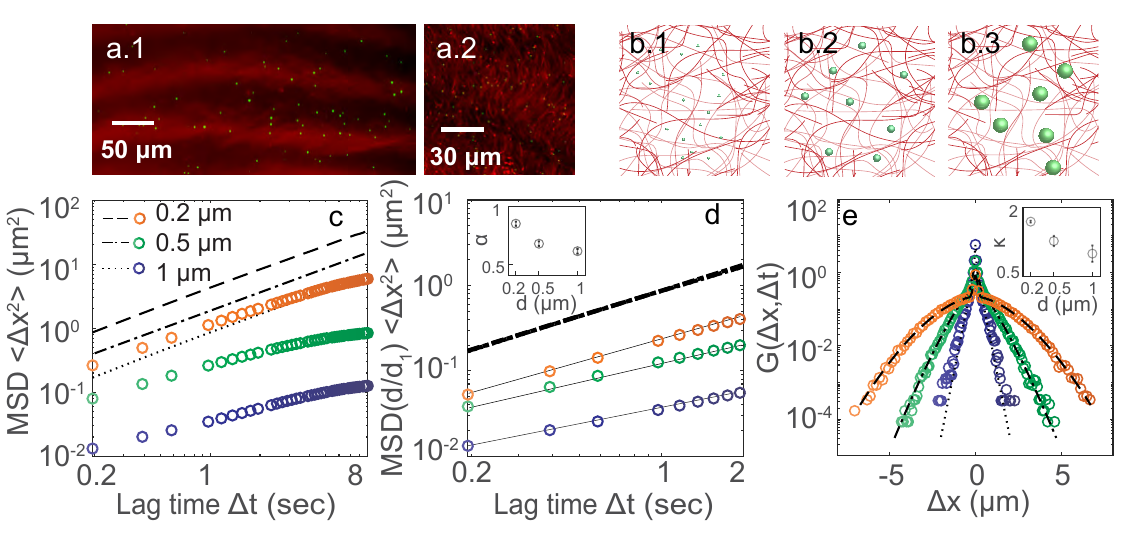}

\caption{Particle mobility is increasingly hindered within the cellulose film as the particle size increases. a.1) Image of the native film from the side interface with \unit{0.5}{\micro\meter} particles embedded inside captured by light-sheet microscopy. a.2) Image of the native film from the top interface with \unit{0.5}{\micro\meter} particles embedded inside captured by confocal microscopy. b.1$-$b.3) Schematics of the particles with different sizes, \unit{0.2}{\micro\meter} (b.1), \unit{0.5}{\micro\meter} (b.2), and \unit{1}{\micro\meter} (b.3) diffusing in the film grown with no alginate (native film).  Ensemble average mean squared displacement, MSD (c), scaled mean squared displacement (d), and distributions of particle displacements at $\Delta$t=2 s (e) of the particles with diameter of \unit{0.2}{\micro\meter} inside the film (orange circle) and in water (dashed line), of \unit{0.5}{\micro\meter} inside the film (green circle) and in water (dash dotted line), and of \unit{1}{\micro\meter} inside the film (purple circle) and in water (dotted line). Inset in (d) plots the sub-diffusive exponent ($\alpha$) as a function of tracer particle diameter and inset in (e) plots the stretching exponent ($\kappa$) as a function of tracer particle diameter.}%
\label{fig:MSDallparticles}%
\end{figure}

We first investigate the effect of 
colloidal tracer size on diffusive 
mobility within a native film. 
A representative 
light-sheet image of tracer 
particles with diameter 
of \unit{0.5}{\micro}{\meter} 
reveals that particles are well-dispersed 
within the porous layers of the 
native film (Figure \ref{fig:MSDallparticles}a.1). 
The embedded tracer particles 
within the films are monodisperse 
and remain monodisperse 
during the experiment though their shape 
can appear a bit distorted by 
the combination of the two beams 
used to create the light-sheet \cite{remacha2020define}, 
Figure \ref{fig:MSDallparticles}a.1. 
We further confirm the tracers 
remain monodisperse in the films 
with confocal microscopy
images in Figure \ref{fig:MSDallparticles}a.2. 
The colloidal particles can only diffuse within the porous layers 
because the pore opening 
of the dense layers is smaller than all of the tracer 
particle sizes used, schematically illustrated in Figures \ref{fig:MSDallparticles}b.1$-$b.3.

The mean-squared displacement (MSD) of the tracers significantly decreases 
as the particle size increases in both the biofilm and in 
water (Figure \ref{fig:MSDallparticles}c), consistent with the 
reduced diffusivity of larger colloids. 
Mobility inside the film, Figure \ref{fig:MSDallparticles}c, symbols, 
is noticeably smaller than in water, Figure \ref{fig:MSDallparticles}c dashed lines, 
for all three tracer sizes because of the physical confinement by the fiber mesh. 
The MSD of particles inside the film grows more slowly, when compared to the free 
diffusion of the particles in water, and plateaus at longer lag times, Figure \ref{fig:MSDallparticles}c.
The power-law exponent of the MSD for all three particles in the film as a 
function of time is smaller than one (\( MSD\sim t^\alpha\), $\alpha<$1), 
indicating confined mobility and sub-diffusive dynamics within the network.

We scaled the MSD of each tracer size by that of the largest 
tracer particle diameter, $d_1 = $ \unit{1}{\micro\meter}, 
to distinguish between the effects of film confinement and tracer 
particle size  on mobility.
The scaling causes perfect 
overlap of the data for tracers in water, 
bold dashed line in Figure \ref{fig:MSDallparticles}d, 
and eliminates variations in mobility due to particle size, 
symbols in Figure \ref{fig:MSDallparticles}d.
 The scaling also indicates the fluid 
in the film has the viscosity of water and any hindered mobility is due to confinement effects.
The scaling also allows us to see how film confinement
affects tracer mobility, Figure \ref{fig:MSDallparticles}d. 
The average pore size of the porous layers of 
the native film is  \unit{0.5 - 1}{\micro}{\meter}, and we can 
use this information to anticipate transport within the network.  

Here, within the time frame of the experiment, 
particles with diameter of \unit{1}{\micro\meter} 
are mostly confined by the network, 
as the local cages never disentangle. 
Smaller particles with diameters of \unit{0.2}{\micro\meter} 
and \unit{0.5}{\micro\meter} are also partially hindered 
due to the network heterogeneity.
Particles can freely diffuse 
within larger pores, but remain partially or 
permanently trapped within the smaller pores 
and exhibit sub-diffusive motion at short time scales. 
Increasing tracer particle diameter then also increases 
the degree of sub-diffusive transport within the 
fiber mesh. 
The sub-diffusive exponent of particles of different sizes, 
estimated from the slope of 
the logarithmic mean-squared displacement, decreases as the 
particle size increases (Figure \ref{fig:MSDallparticles}d, inset), 
and particles slow down as a result of 
increased encounters with bounding fibers (Table S2). 
Similar to our observation, confined mobility of particles 
in a cross-linked polymeric network is sub-diffusive ($\alpha=0.5$) 
for time lags smaller than the relaxation time of a polymer 
chain \cite{cai2015hopping}. At larger time scales, particles are trapped and the MSD approaches a plateau. 

The distribution of particle displacements measured at  $\Delta t= 2$ s 
is plotted in Figure \ref{fig:MSDallparticles}e.
Two distinct populations are noted for all particle sizes. 
One population, at the center of the distribution near $\Delta x = 0$, 
can be attributed to nearly immobile particles.
The second, broader population is the mobile or partially mobile particles at 
higher magnitude $\Delta x$ values, Figure \ref{fig:MSDallparticles}e. 
The different populations can be described 
by fitting the distribution of particle displacements 
to a sum of a Gaussian function and a stretched exponential 
function \cite{valentine2001investigating, Chaudhuri2007} 
expressed as:
\[ G_s(\Delta x, \Delta t) = a_1 \exp \bigg(- \left( \frac {\Delta x} {\delta}\right)^2\bigg) + a_2 \exp \bigg(- \Big| \frac{\Delta x} {\gamma (\Delta t)}\Big|^{\kappa} \bigg) \]

The Gaussian function fits the center of the distribution, 
quantifying the trapping of some 
particles within very small pores of the heterogeneous film. 
The stretched exponential tail of the distribution, 
on the other hand, characterizes 
the particles that are moving within the pore 
space and are only partially hindered by 
pore wall confinement.
The stretched exponential fit provides a value 
of the stretching exponent, $\kappa$, that 
quantifies the confinement effects on mobility and 
deviations from Gaussian particle diffusion. 
As the particle size increases, the stretching exponent 
decreases as we see in Figure \ref{fig:MSDallparticles}e, inset.  
The stretching exponent of the particles with diameter 
of \unit{0.2}{\micro\meter} is close to the Gaussian 
value, $\kappa = 2$, while the value for the 
particles with diameter of \unit{1}{\micro\meter} is 
close to $\kappa = 1$ (Figure \ref{fig:MSDallparticles}e, inset), 
indicating deviation from Gaussian dynamics for the larger tracer particles. 
The dynamics of the mobile, and partially mobile, 
particles are also governed by particle size. 
For particle diameters more than five times smaller than 
the average pore size, movement is not confined within the voids.

Dynamical heterogeneity in other soft glassy systems leads to a  
non-Gaussian distribution of displacements, where the 
motion is decoupled into fast and slow populations during 
the structural 
relaxation \cite{Chaudhuri2007,Gao2007,roberts2019tracer}. 
Entrapped particles within 
the structure form a Gaussian center and free particles 
form an exponential tail of the distribution, just as 
we see here. 
Diffusion of particles within actin filament solutions \cite{Wang2012,Stuhrmann2012}, 
through cells \cite{Lampo2017}, and within colloidal suspensions \cite{kegel2000direct,Gao2007} 
is also observed to be non-Gaussian due to the 
spatiotemporal heterogeneity of the environment,
where the elasticity of the structure controls the 
behaviour of the sub-populations. Non-Gaussian behaviour 
is observed because individual particles have heterogeneous 
dynamics temporally along their movements and spatially 
compared to other particles, where series of Gaussian events 
with changing variance lead to observation of exponential 
behavior \cite{wang2009anomalous}. 

Here, subdivision of the particle displacements into the slow and fast 
dynamics mostly reflects the spatial heterogeneity of the 
underlying fiber mesh structure, producing 
clear non-Gaussian behaviour. The heterogeneous mobility of 
particles trapped within pores of different sizes creates a 
series of Gaussian distributions with different variance for 
the mobile populations forming the tail. The collective 
distribution of all the events is therefore non-Gaussian 
\cite{wang2009anomalous} and the stretching exponent shows 
the degree of variability of the distributions of the mobile 
population. 

\begin{figure}
\center
\includegraphics[scale=0.90]{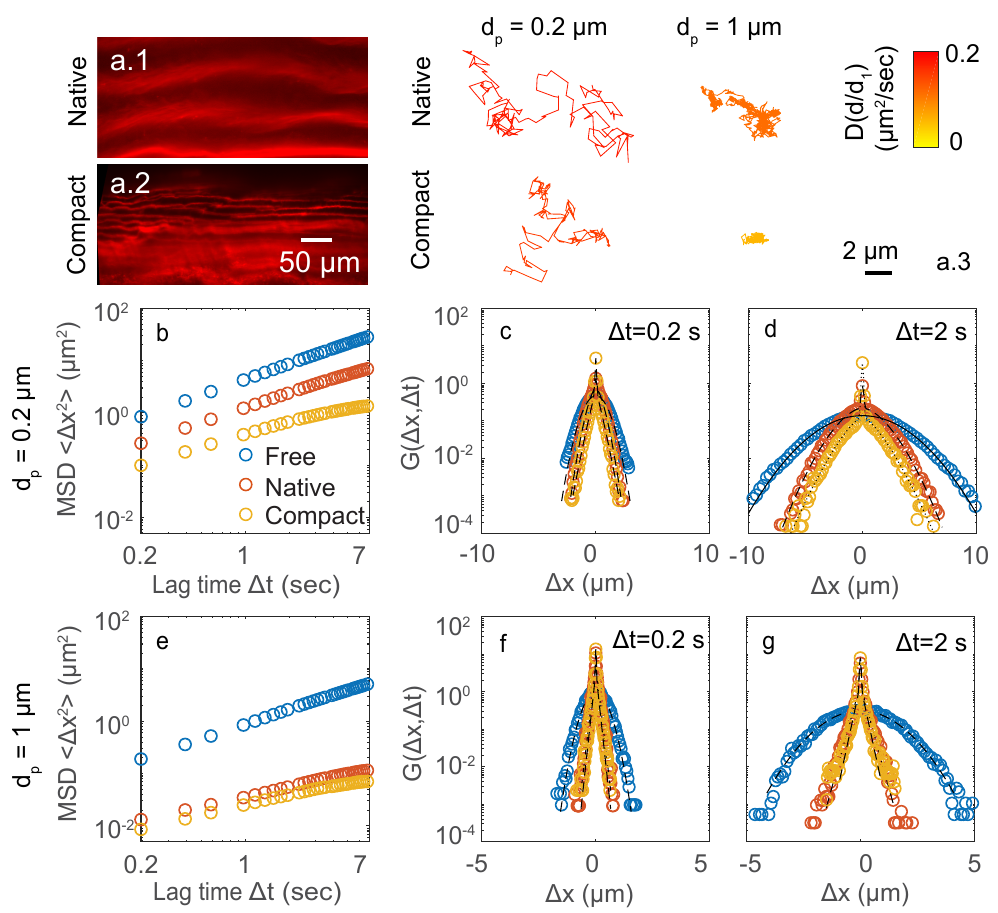}

\caption{Tracers dynamics vary in different spatial structures. Micro-structure of native (a.1) and compact (a.2) films. Representative trajectories (a.3) of the particles of different sizes (0.2 and \unit{1}{\micro\meter}) diffusing in the native (a.1) and the compact (a.2) films. The color shows the normalized average diffusion coefficient of the trajectories. b) Mean squared displacement, c) distributions of particle displacements at $\Delta$t=0.2 s and d) at $\Delta$t=2 s of particles of \unit{0.2}{\micro\meter} diffusing in water (blue circle), in the native films (red circle), and in the compact film (yellow circle). e) Mean squared displacement, f) distributions of particle displacements at $\Delta$t=0.2 s and g) at $\Delta$t=2 s of particles of \unit{1}{\micro\meter} diffusing in water (blue circle), in the native film (red circle), and in the compact film (yellow circle). 
}%
\label{fig:MSD_allbiofilms}%
\end{figure}

Dynamics of the tracer particles with diameters of \unit{0.2}{\micro\meter} 
and \unit{1}{\micro\meter} are next compared in  
two films with different overall fiber mesh densities. 
The native film has a bulk fiber 
density of \unit{0.01}{\gram\per\centi\cubic\meter} 
while the compact film has a density of \unit{0.03}{\gram\per\centi\cubic\meter}.
A light-sheet micrograph of the native film is shown 
in Figure \ref{fig:MSD_allbiofilms}a.1, formed with no alginate present, and 
a more compact film in Figure \ref{fig:MSD_allbiofilms}a.2 
formed with $1$ \% $w/v$ of sodium alginate present. 
Trajectories of the \unit{0.2}{\micro\meter} 
and \unit{1}{\micro\meter} exhibit a 
pronounced difference in the native (Figure \ref{fig:MSD_allbiofilms}a.3, top row) 
and the compact (Figure \ref{fig:MSD_allbiofilms}a.3, bottom row) biofilms. 
To eliminate the effect of particle size, particle trajectories 
are colored based on the value of their normalized diffusivity, $D(d/d_1)$. 
Trajectories of particles with diameter of \unit{0.2}{\micro\meter} are 
essentially free to diffuse with only intermittent hindrance. 
Particles with diameter of \unit{1}{\micro\meter}, however, 
follow trajectories with longer intervals of slowed mobility. 
The representative trajectories show reduced diffusivity when the 
particle size increases or when the density of the network increases. 
To quantify the effect of mesh density on tracer mobility, the MSD and 
the distribution of particle displacements are compared for mobility of 
particles within different films.

The MSD of the tracers with diameter of  \unit{0.2}{\micro\meter} 
in the native film is about five times smaller than for diffusion in water. 
Diffusion in the alginate-grown compact film is also reduced by a factor of two 
compared to the native film and by a factor of 10 compared to free aqueous diffusion (Table S2). 
The slope of the logarithmic MSD at short times ($\Delta t < 5$ s) 
decreases from 1 in water to $\sim 0.9$ in the native film and to 
$\sim 0.7$ in the compact film (Figure \ref{fig:MSD_allbiofilms}b). 
The difference in the microstructure of the two biofilms (Figure \ref{fig:biofilms} c and d) 
can be assessed based on the degree of hindrance of particles 
with the same size in each film. Reduced MSD and increased 
sub-diffusivity of \unit{0.2}{\micro\meter} particles 
in the compact film indicates the local pores, on average, 
confine particles to a greater extent.

The distribution of particle displacements, represented at 
$\Delta t$ = 0.2 and 2 s (Figure \ref{fig:MSD_allbiofilms} c $-$ d), 
also signifies the difference between hindered mobility within 
the native and compact films. The distribution of particle 
displacements exhibits a slightly narrower tail within the 
films
compared to the distribution within water and also 
forms two populations of displacements. 
The difference is more visible at 
longer time lags ($\Delta t$ = 2 s), when 
the average particle 
displacement increases and particles 
more frequently encounter the fiber 
mesh (Figure \ref{fig:MSD_allbiofilms} d and g). 
Here, the tails of the distributions are significantly narrower 
within the films, 
indicating reduced mobility at 
longer time lags. The difference in 
the behavior of the 
mobile population of  particles 
within the two films 
reveals that even local displacements are controlled by the 
ratio of the particle to void diameter (Figure \ref{fig:MSD_allbiofilms} d and g). 

The MSD of \unit{1}{\micro\meter} tracers decreases 
by more than one order of magnitude
in both film types, compared to the 
dynamics in water, 
and is slightly smaller 
in the compact than the native 
film (Figure \ref{fig:MSD_allbiofilms}e). 
The slope of the logarithmic MSD 
is also reduced from 1 in water to $\sim 0.6$ in the native and compact films. 
Similar to what is observed in the dynamics of \unit{0.2}{\micro\meter} tracers, 
the width of the distribution of particle displacements with diameter 
of \unit{1}{\micro\meter} reduces within the two films and the 
difference is more obvious at longer time lags. The distributions of 
particle displacements within the two films, however, nearly overlap 
at short and long time lags. The mobility of the \unit{1}{\micro\meter} 
particles is strictly hindered within the films, where the particle 
diameter is close to the average mesh pore size and the dynamics 
are not significantly affected by the mesh heterogeneity (Figure \ref{fig:MSD_allbiofilms} f and g), 
because both are confining. 
The dynamics of the \unit{0.2}{\micro\meter} tracers seem 
to be more sensitive to the change in the structure of the films and 
the confinement effect on the reduced mobility of the \unit{0.2}{\micro\meter} 
tracers in the compact film is much greater. The temporal trajectories of single-particle displacements (Figure S1) show that mobile particles have random displacement till they disappear from the imaging volume. The average residence time of the particles increases as particle size increases. The immobile particles remain mostly stationary throughout the imaging sample time. However, a small portion of the immobile particles ($\approx$ 5\%) infrequently jump from their cages. The waiting time between the jumps depends on the ratio of the particle to the cage size and spans from short to long times because of the pore heterogeneity.

\begin{figure}
\center
\includegraphics[scale=0.85]{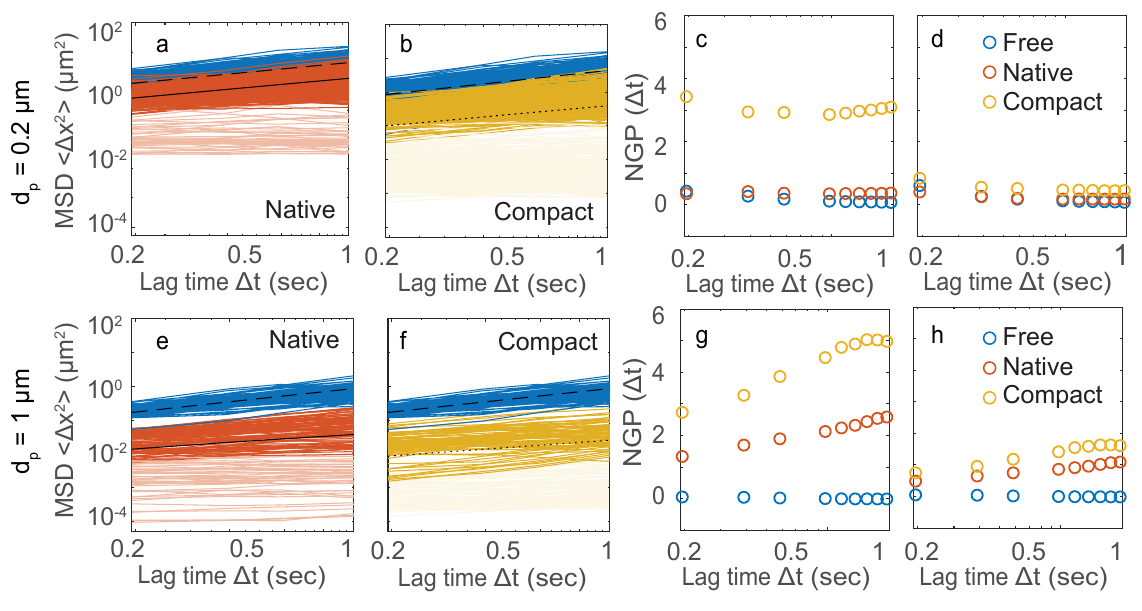}

\caption{Individual tracers experience different levels of confinement within the heterogeneous films. Mean squared  displacement of single  \unit{0.2}{\micro\meter} particles diffusing in the native film, classified in two groups of mobile (dark red lines in a) and immobile (light red lines in a) and in the compact film, classified in two groups of mobile (dark yellow lines in b) and immobile (light yellow lines in b). Blue lines in a and b show free diffusion of \unit{0.2}{\micro\meter} tracers and black dashed lines show the ensemble MSD. Non-Gaussian parameters of all tracers (c) and only mobile  \unit{0.2}{\micro\meter} tracers (d) diffusing in water (blue circle), in the native films (red circle), and in the compact films (yellow circle). Mean squared  displacement of single \unit{1}{\micro\meter} tracers diffusing in the native biofilm, classified into groups of mobile (dark red lines in e) and immobile (light red lines in e) and in the compact biofilm, classified into groups of mobile (dark yellow lines in f) and immobile (light yellow lines in f). Blue lines in e and f show free diffusion of \unit{1}{\micro\meter} tracers in water and black dashed lines show the ensemble MSD. Non-Gaussian parameters of all tracers (g) and (h) only mobile \unit{1}{\micro\meter} tracers diffusing in water (blue circle) and in the native (red circle), and in the compact (yellow circle) films. Black lines in a and e show the ensemble MSD of \unit{0.2}{\micro\meter} and \unit{1}{\micro\meter} tracers in the native film and black dotted lines in b and f show the ensemble MSD of \unit{0.2}{\micro\meter} and \unit{1}{\micro\meter} tracers in the compact film. }%
\label{fig:SinlgeMSD_allbiofilms}%
\end{figure}

 We calculate the MSD of single \unit{0.2}{\micro\meter} 
 and \unit{1}{\micro\meter} 
 tracers to obtain more insight into individual tracer size effects 
 on movement within the films. The logarithmic MSD of the 
 individual particles in both native and compact films 
 is more widely distributed than the MSD of single particles 
 in water (Figure \ref{fig:SinlgeMSD_allbiofilms}a, b, e, and f). 
 We classify the individual MSDs with large varieties into two 
 categories of immobile and mobile, using  
 thresholds of 0.1 $\mu m^2 $ for the \unit{0.2}{\micro\meter} 
 tracers and 0.01 $\mu m^2 $ for the \unit{1}{\micro\meter} 
 tracers, based on the crossover point between the two 
 populations at $\Delta t = 2$ s. (Figure \ref{fig:MSDallparticles}d and g). 
 
 The categorized MSDs illustrate the effect of the 
 underlying biofilm structure on the local mobility of 
 single tracer particles and the two film types show distinct behavior 
 for the \unit{0.2}{\micro\meter} tracers. 
 The single MSDs of the mobile category 
 have greater overlap with the MSDs of the particles 
 in water (Figure \ref{fig:SinlgeMSD_allbiofilms}a), 
 indicating the presence of more particles with 
 unconfined motions in the native film. 
 The immobile category of the particles within 
 the compact film has larger population with 
 smaller mobility compared to the immobile category in 
 the native film. The structure of the native film has, 
 on average, less confining effect than the compact 
 film on the mobility 
 of the \unit{0.2}{\micro\meter} tracers 
 (Figure \ref{fig:SinlgeMSD_allbiofilms}a and b). 
Increased tracer size removes these distinctions, 
as the mobile and immobile categories of the 1 $\mu m$ 
 tracers are very similar in the native and the compact films. 
 The mobility of the mobile or partially mobile category of all tracer 
 trajectories is smaller than the mobility of particles in water. 
 (Figure \ref{fig:SinlgeMSD_allbiofilms}e and f). 
 The observations indicate the compact film's small pores limit 
 the mobility of both \unit{0.2}{\micro\meter} and 
 \unit{1}{\micro\meter} tracers, meaning we can control 
 biofilm transport rates across length scales relevant to 
 colloidal and bacterial diffusion.

The non-Gaussian parameter \cite{weeks2000three}, \( \alpha (\Delta t) = ((<\Delta x^4>)/(3<\Delta x^2>^2)-1)\), over time quantifies the effects of 
confinement on the immobile tracers identified earlier. 
The $\alpha$ of the total population of 
 \unit{0.2}{\micro\meter} tracers is slightly 
greater than zero in the native film, but notably larger 
in the compact film (Figure \ref{fig:SinlgeMSD_allbiofilms}c) and 
 increases over time as mobility becomes more hindered. 
For the mobile sub-population of the \unit{0.2}{\micro\meter} 
particles, however, the non-Gaussian parameter is closer 
to zero within the native and the compact film. Removing the 
immobile population from the displacements of 
the \unit{0.2}{\micro\meter} tracers reduces 
the non-Gaussian parameter to 
near zero for all times (Figure \ref{fig:SinlgeMSD_allbiofilms}d). 
Dynamic heterogeneity, however, produces a measurable 
non-Gaussian parameter for the mobile sub-population 
of \unit{0.2}{\micro\meter} tracers within the compact film. 

The estimated non-Gaussian parameter of the 
\unit{1}{\micro\meter} tracers is non-zero 
in both native and compact films for the 
total tracer population and is larger in the 
compact film 
(Figure \ref{fig:SinlgeMSD_allbiofilms}g), 
quantifying the effects of 
the two films' different pore size distributions. 
Here removing the immobile sub-population 
reduces the non-Gaussian parameter, indicating 
the mobile population diffusion is also quite non-Gaussian 
in the biofilms (Figure \ref{fig:SinlgeMSD_allbiofilms}h).
Even though the \unit{1}{\micro\meter} tracers are mobile within the two films, their movements are 
heterogeneous and non-Gaussian.

Quantifying the percentage of particles that are immobile and mobile provides insight to differentiate the pore characteristics of the network. The immobile tracers are hindered in their motion by pores that are similar in size to the particles. This explanation is supported by the fact that the percentages of immobilized tracers increase as the size of the pore and the tracer particle change and become more similar. Comparing the percentage of the mobile and the immobile particles of different sizes in the native film shows that as particle size increases, the percentage of immobile particles within the film increases. For the 0.2 $\mu m$ tracers within the native film,  only 5\% of the particles are immobile, while the percentage of immobile 0.5 and 1 $\mu m$ particles increases to 10\% and 37\%, respectively, within the native film. This indicates the existence of small pores with diameter of 0.5 – 1 $\mu m$, limiting the mobility of particles with a size greater than 0.5 $\mu m$. The percentage of immobile particles in the compact film is larger compared to the percentage in the native film. The difference is more significant for the 0.2 and 0.5 $\mu m$ tracers and is 29\% and 23\%, respectively. For the 1  $\mu m$ particles, half (49\%) remain mostly immobile. The compact film, on average, thus has smaller pores, about 0.2 – 0.5 $\mu m$, compared to the native film, and hinders the mobility of particles with smaller sizes.

\begin{figure}
\center
\includegraphics[scale=0.98]{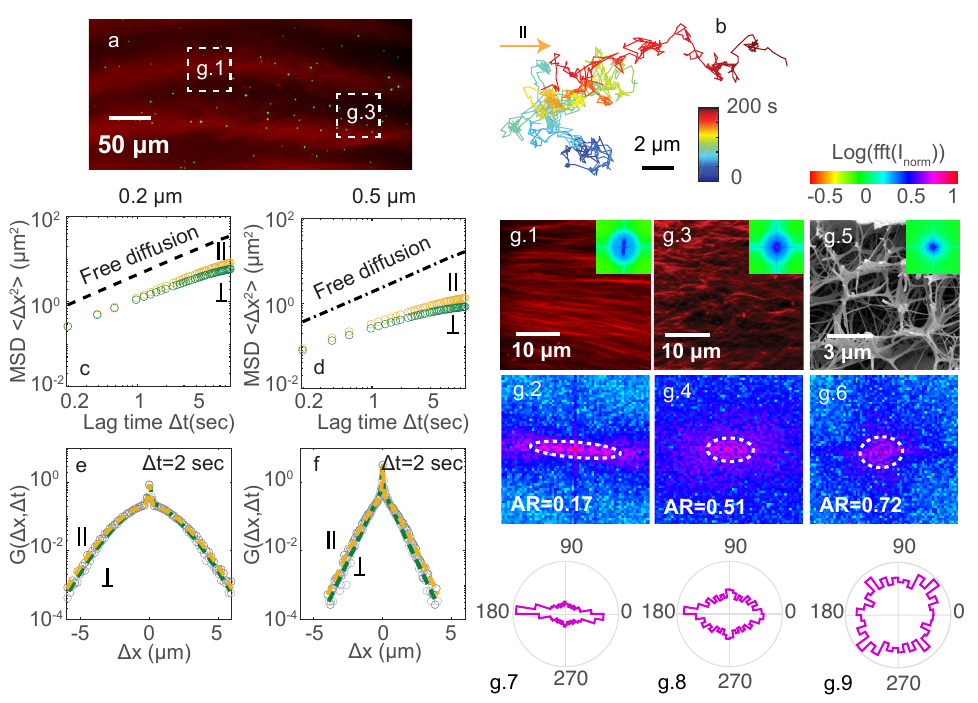}
\caption{The structural orientation of the film influences dynamics of the tracer particles in the x and y directions. a) Structure of the native film from the side interface with \unit{0.5}{\micro\meter} diameter particles diffusing in the layers. b) A representative trajectory of a \unit{0.5}{\micro\meter} particle in the native film. The color bar represents the diffusion time of the particle. Ensemble average mean squared displacements of (c) \unit{0.2}{\micro\meter} and (d) \unit{0.5}{\micro\meter} particles in the horizontal and vertical directions in the film. Distribution of particle displacements of (e) \unit{0.2}{\micro\meter} and (f) \unit{0.5}{\micro\meter} particles at $\Delta t=2 $ sec for horizontal and vertical mobility.
The magnified confocal images from the layer (g.1) the interlayer locations (g.3) and the SEM image of the interlayer location (g.5). The inset represents 2D-FFT images of each location. The magnified images of the low spatial frequencies shifted by 90 degree for the confocal-layer (g.2) and the confocal-interlayer (g.4) and the SEM-interlayer (g.6). The polar distribution of radially integrated angles for the layer (g.7) and the interlayer (g.8) and the SEM-interlayer (g.9) depict degree of alignment of the fibers in the horizontal directions. Color bar of the 2D-FFT images represents logarithmic power spectrum values of the normalized intensity of the images.}%
\label{fig:par_perp}%
\end{figure}

Previous work \cite{Stylianopoulos2010} has identified 
strongly anisotropic diffusive mobility in fiber matrices 
and the film studied here might be expected 
to exhibit similar effects. 
We assess diffusive mobility of \unit{0.2}{\micro\meter} 
and \unit{0.5}{\micro\meter} particles in the 
horizontal/parallel and vertical/perpendicular directions of the 
native film as a means of quantifying 
any anisotropic transport. 
Micron-sized particles can only diffuse in 
the low-density regions of the films between the 
dense layers so those results are the focus here 
(Figure \ref{fig:par_perp}a). 

The long-time trajectories of particles exhibit 
a more horizontally dominated movement, parallel to  
the typical void orientation (Figure \ref{fig:par_perp}b). 
The MSD of the \unit{0.2}{\micro\meter} particles 
is 1.4$\times$ greater in the parallel direction 
than the perpendicular direction, (Figure \ref{fig:par_perp}c). 
The difference is more evident in the MSD 
of \unit{0.5}{\micro\meter} particles 
(Figure \ref{fig:par_perp}d). 
The mobility of the larger \unit{0.5}{\micro\meter} 
particles is more limited so the effect 
of structural orientation of the media on 
mobility of the particles is stronger, 
1.6$\times$ (Figure \ref{fig:par_perp}d). 

The distribution of the particle displacements also 
confirms that particles tend to diffuse slightly faster 
through the direction parallel to the 
horizontally-oriented pores (Figure \ref{fig:par_perp}e and f). 
The biofilm is formed and thickened at the 
air-liquid interface, where bacteria are 
largely constrained to two-dimensional, 
lateral movement. As a result, the intermediate 
densified layers are formed from a 
horizontally-aligned network of fibers 
(Figure \ref{fig:par_perp}g.1). The porous layers 
between the dense layers have more random 
fiber orientation, (Figure \ref{fig:par_perp}g.3 and g.5). 

A 2D fast Fourier transform (2D-FFT) analysis 
of the confocal images of the dense and porous 
layers, inset of \ref{fig:par_perp}g.1 and g.3, 
and the SEM image of the porous layer, 
inset of \ref{fig:par_perp}g.5, quantify 
the structural anisotropy. The magnified low 
spatial frequencies of the 2D-FFT images, 
shifted by 90$\degree$, are aligned in the 
parallel direction (Figure \ref{fig:par_perp}g.2, g.4, and g.6). 
The center of the power spectrum is fitted 
to an ellipse and the ratio of the minor 
to major axis length is defined as the 
directionality aspect ratio, AR \cite{ghazaryan2012analysis}. 
The smaller the AR, the more fibers are aligned, 
with the dense layer exhibiting an $AR=0.17$. 
Confinement anisotropy reduces significantly within 
the porous layer with $AR=0.72$. 
Distribution of the fiber angle orientation 
is calculated from the radially-integrated 
value of the 90$\degree$-shifted power spectrum 
at each angle (Figure \ref{fig:par_perp}g.7 $-$ g.9). 
The polar distribution of angles, in agreement 
with the estimated AR, indicates 
a reduced confinement anisotropy 
for the fibers in the porous layers. 
Particle tracking at different 
length scales also confirms that, although mobility is 
slightly hindered in the perpendicular 
direction, confinement anisotropy 
within the porous region is minimal. 

\begin{figure}
\center
\includegraphics[scale=1]{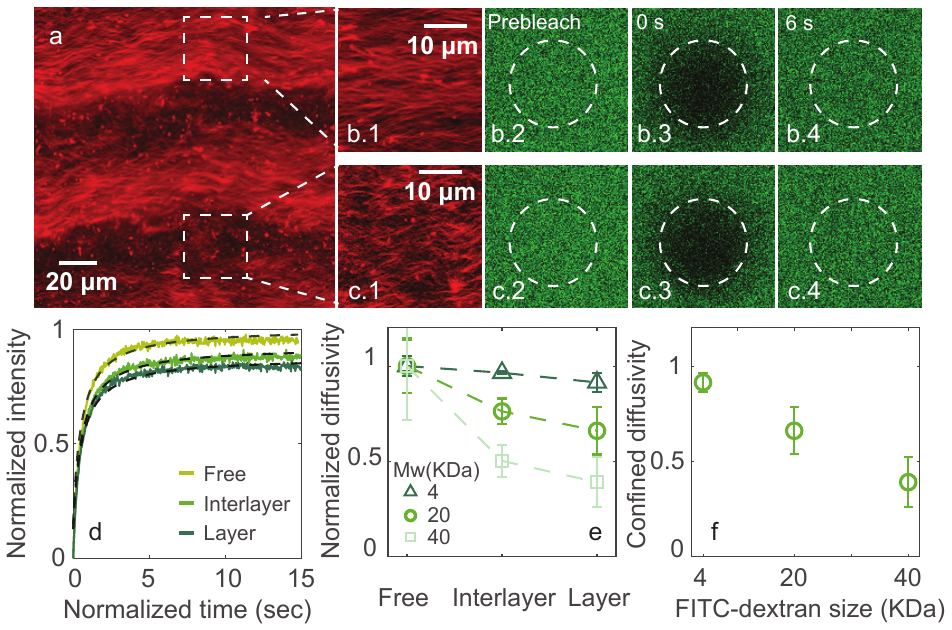}

\caption{Molecular mobility is reduced as FITC-dextran diffuses through the biofilm. a) Confocal image of the biofilm side interface highlights the layered microstructure of the film. b,c) Zoomed-in images of the structure in a (b.1) porous and (c.1) dense layer. Intensity of FITC-dextran (20 kDa)  (b.2 and c.2) prior to bleaching, (b.3 and c.3) just after bleaching, and  (b.4 and c.4) after recovery at t = 6 s for selected locations inside and between dense layers. d) Normalized intensity of FITC-dextran (20 kDa) over time for recovery after bleaching within water (light green), the porous layer (medium green) and the dense layer (dark green) with the fitted model. e) Average normalized diffusivity ($D/D_{free}$) of FITC-dextran with different molecular weights (4 - 40 kDa) in different areas. f) Confined normalized diffusivity ($D/D_{free}$) within the dense layers for FITC-dextran of different molecular weights.}%
\label{fig:frap}%
\end{figure}

The periodic structure of the biofilms 
results from the regular formation of dense, more 
consolidated layers of fibers with 
an average 2D fiber fraction of 
$64 \pm 5 \%$, measured from the 
confocal images (Figure \ref{fig:frap} a and b.1). 
The porous layers between the dense layers have a  
lower fiber area density of $42 \pm 4 \%$ (Figure \ref{fig:frap} a and c.1). Tracer 
particles with diameters between \unit{0.2}{\micro\meter} and \unit{1}{\micro\meter} can only 
diffuse within the porous layers and 
can not penetrate the dense 
layers of either film, Figure \ref{fig:par_perp}a, 
though macromolecules can. 

The barrier properties of the dense layers 
are assessed by measuring molecular 
diffusion of fluorescein 
isothiocyanate-dextran (FITC$-$dextran) 
with molecular weights between 4 and 
40 kDa using Fluorescence Recovery 
After Photobleaching (FRAP). 
Selected areas inside, 
Figure \ref{fig:frap} b.2 $-$ b.4, and 
between, Figure \ref{fig:frap} c.2 $-$ c.4, 
the dense layers were studied to provide an overall 
characterization of the composite film. 
During recovery, fluorescent dye diffused back into 
the circular bleached region, and the 
normalized fluorescent intensity is 
calculated as \(I_{norm}(t)=\frac {I(t)-I_{bleach}} {I_{pre}-I_{bleach}} \),
where $I_{norm}(t)$ is the normalized intensity, 
$I(t)$ is the intensity of the bleached 
area over time during recovery, $I_{bleach}$ 
is the average bleached intensity, 
and $I_{pre}$ is the average intensity 
before bleaching. The recovery of the 
normalized intensity is fitted to \(I_{norm}(t)=A.e^{-2\tau_D/t}\Bigg(I_0\Big(2\tau_D/t\Big)+I_1\Big(2\tau_D/t\Big)\Bigg)\) assuming 
lateral diffusion within the film, 
where $A$ is the pre-exponential factor 
and $\tau_D$ is the half time recovery 
of the normalized intensity \cite{soumpasis1983theoretical}. 
The diffusion coefficient can be estimated as \(D=0.88w^2/{4\tau_D}\) for a circular bleached area with a radius of $w$ \cite{carnell2015fluorescence}. 
Diffusive mobility of 4 kDa FITC$-$dextran 
is not hindered within the porous 
and the dense layers of the biofilm 
compared to the free diffusive mobility. 
Increasing the FITC$-$dextran molecular 
weight to 20 and 40 kDa, however, 
reduces the average diffusivity. 

The diffusive dynamics are more 
confined (reduced by $\sim$ 25 \%) 
inside the dense layers, Figure \ref{fig:frap}e, 
and the confined mobility, on average, 
decreases as the FITC$-$dextran size 
increases, Figure \ref{fig:frap}f 
and Table S3. 
The behavior is consistent with the 
observed hindered dynamics of micron-size 
tracer particles in the film. The 
heterogeneous structure of the biofilm 
contains a wide distribution of pore sizes. 
While the average pore size of the network 
(\unit{0.5-1}{\micro\meter}) is 
much larger than the size of the 
FITC$-$dextran molecules, the 
molecular mobility is still hindered 2$\times$, 
meaning that network connectivity and 
heterogeneity can also affect 
the molecular diffusion in the film. 
Similar confined mobility was 
previously measured for protein 
secretion and diffusion of 
macromolecules within hydrogel 
networks \cite{Hsu2018,Pedron2015,Deforest2015,Loebel2020}. 

\section*{Conclusions}
Transport within bacterial cellulose biofilms 
is a phenomenon central to many applications of 
the unique fiber mesh material, as well as 
design and simulation of these materials, whether the 
goal is diffusion of nutrients, cells, or ions.
We might naively assume these films have an 
entirely random 
fiber structure, but show using microscopy that 
the films have periodic structures 
of very dense oriented fiber layers alternating with 
more porous random fiber layers.  
Particle tracking provides an excellent 
measure of the complex, 
anisotropic pore structures 
within the fiber network 
as an indicator of specific 
performance parameters like mesh size, 
pore diameter, and permeability. 
Complementary confocal and light-sheet 
microscopy techniques enable assessment 
of the effects of process fermentation 
conditions and provide a means of 
specifying the performance needed for 
various applications.

The spatially varying packing 
density of the bacterial cellulose biofilms 
presents a complex  pathway 
for diffusing particles, where the 
relative tracer particle diameter and 
network structure control the mobility. 
As these dimensions can radically alter 
accessibility of the film by entities of certain sizes,
we demonstrate that the overall compactness of the 
film can be controlled during its 
growth by adding sodium alginate 
to thicken the culture medium. 

The MSD of particles with varying sizes 
is sub-diffusive and non-Gaussian 
in the biofilms, with the level of 
sub-diffusivity increasing as the 
particle size increases or the 
compactness of the network increases. 
Consequently, tracer trajectories vary 
significantly depending on relative dimensions.
Mobility at different length scales in 
the film is slightly more hindered in 
the direction perpendicular to the 
predominant fiber growth direction. 
Distinctive dynamics of the particles 
in different media can be utilized 
to probe the underlying structure 
of the biofilms. Understanding the 
pronounced structural variation of 
complex fibrous networks over micron 
length scales could enhance the accuracy of 
efforts to design synthetic tissue scaffolds, 
for example, or improve 
modeling\cite{Stylianopoulos2010} 
of transport of nutrients or 
antibiotics within fibrous biofilms.

\begin{suppinfo}
MSD cutoff; Temporal trajectories; Diffusivity of tracers and macromolecules within the native film and the compact film; Diffusivity of tracers and macromolecules within the native film and the compact film

\end{suppinfo}

\begin{acknowledgement}
Light-sheet and confocal microscopy and scanning electron microscopy were performed using instruments 
situated in, and maintained by, the Katharina Gaus Light Microscopy Facility (KGLMF) and Electron Microscope 
Unit (EMU) at the Mark Wainwright Analytical Centre, UNSW Sydney. 
MH acknowledges a UNSW Scientia Fellowship. Partial support from ARC DP190102614 is gratefully acknowledged. 
\end{acknowledgement}

\bibliography{Diffusion_in_biofilm}

\providecommand{\latin}[1]{#1}
\makeatletter
\providecommand{\doi}
  {\begingroup\let\do\@makeother\dospecials
  \catcode`\{=1 \catcode`\}=2 \doi@aux}
\providecommand{\doi@aux}[1]{\endgroup\texttt{#1}}
\makeatother
\providecommand*\mcitethebibliography{\thebibliography}
\csname @ifundefined\endcsname{endmcitethebibliography}
  {\let\endmcitethebibliography\endthebibliography}{}
\begin{mcitethebibliography}{75}
\providecommand*\natexlab[1]{#1}
\providecommand*\mciteSetBstSublistMode[1]{}
\providecommand*\mciteSetBstMaxWidthForm[2]{}
\providecommand*\mciteBstWouldAddEndPuncttrue
  {\def\EndOfBibitem{\unskip.}}
\providecommand*\mciteBstWouldAddEndPunctfalse
  {\let\EndOfBibitem\relax}
\providecommand*\mciteSetBstMidEndSepPunct[3]{}
\providecommand*\mciteSetBstSublistLabelBeginEnd[3]{}
\providecommand*\EndOfBibitem{}
\mciteSetBstSublistMode{f}
\mciteSetBstMaxWidthForm{subitem}{(\alph{mcitesubitemcount})}
\mciteSetBstSublistLabelBeginEnd
  {\mcitemaxwidthsubitemform\space}
  {\relax}
  {\relax}

\bibitem[Soares \latin{et~al.}(2021)Soares, de~Lima, and
  Schmidt]{soares2021technological}
Soares,~M.~G.; de~Lima,~M.; Schmidt,~V. C.~R. Technological aspects of
  kombucha, its applications and the symbiotic culture (SCOBY), and extraction
  of compounds of interest: A literature review. \emph{Trends Food Sci.
  Technol.} \textbf{2021}, \emph{110}, 539--550\relax
\mciteBstWouldAddEndPuncttrue
\mciteSetBstMidEndSepPunct{\mcitedefaultmidpunct}
{\mcitedefaultendpunct}{\mcitedefaultseppunct}\relax
\EndOfBibitem
\bibitem[Chakravorty \latin{et~al.}(2016)Chakravorty, Bhattacharya,
  Chatzinotas, Chakraborty, Bhattacharya, and Gachhui]{chakravorty2016}
Chakravorty,~S.; Bhattacharya,~S.; Chatzinotas,~A.; Chakraborty,~W.;
  Bhattacharya,~D.; Gachhui,~R. Kombucha tea fermentation: Microbial and
  biochemical dynamics. \emph{Int. J. Food Microbiol.} \textbf{2016},
  \emph{220}, 63--72\relax
\mciteBstWouldAddEndPuncttrue
\mciteSetBstMidEndSepPunct{\mcitedefaultmidpunct}
{\mcitedefaultendpunct}{\mcitedefaultseppunct}\relax
\EndOfBibitem
\bibitem[Rooney \latin{et~al.}(2020)Rooney, Amos, Hoskisson, and
  McConnell]{rooney2020}
Rooney,~L.~M.; Amos,~W.~B.; Hoskisson,~P.~A.; McConnell,~G. Intra-colony
  channels in E. coli function as a nutrient uptake system. \emph{ISME J.}
  \textbf{2020}, \emph{14}, 2461--2473\relax
\mciteBstWouldAddEndPuncttrue
\mciteSetBstMidEndSepPunct{\mcitedefaultmidpunct}
{\mcitedefaultendpunct}{\mcitedefaultseppunct}\relax
\EndOfBibitem
\bibitem[Tanpichai \latin{et~al.}(2012)Tanpichai, Quero, Nogi, Yano, Young,
  Lindstr{\"{o}}m, Sampson, and Eichhorn]{Tanpichai2012}
Tanpichai,~S.; Quero,~F.; Nogi,~M.; Yano,~H.; Young,~R.~J.;
  Lindstr{\"{o}}m,~T.; Sampson,~W.~W.; Eichhorn,~S.~J. {Effective young's
  modulus of bacterial and microfibrillated cellulose fibrils in fibrous
  networks}. \emph{Biomacromolecules} \textbf{2012}, \emph{13},
  1340--1349\relax
\mciteBstWouldAddEndPuncttrue
\mciteSetBstMidEndSepPunct{\mcitedefaultmidpunct}
{\mcitedefaultendpunct}{\mcitedefaultseppunct}\relax
\EndOfBibitem
\bibitem[Shi \latin{et~al.}(2014)Shi, Zhang, Phillips, and Yang]{shi2014}
Shi,~Z.; Zhang,~Y.; Phillips,~G.~O.; Yang,~G. Utilization of bacterial
  cellulose in food. \emph{Food hydrocolloids} \textbf{2014}, \emph{35},
  539--545\relax
\mciteBstWouldAddEndPuncttrue
\mciteSetBstMidEndSepPunct{\mcitedefaultmidpunct}
{\mcitedefaultendpunct}{\mcitedefaultseppunct}\relax
\EndOfBibitem
\bibitem[Bangasser \latin{et~al.}(2017)Bangasser, Shamsan, Chan, Opoku,
  T{\"u}zel, Schlichtmann, Kasim, Fuller, McCullough, Rosenfeld, \latin{et~al.}
  others]{bangasser2017}
Bangasser,~B.~L.; Shamsan,~G.~A.; Chan,~C.~E.; Opoku,~K.~N.; T{\"u}zel,~E.;
  Schlichtmann,~B.~W.; Kasim,~J.~A.; Fuller,~B.~J.; McCullough,~B.~R.;
  Rosenfeld,~S.~S., \latin{et~al.}  Shifting the optimal stiffness for cell
  migration. \emph{Nat. Commun.} \textbf{2017}, \emph{8}, 1--10\relax
\mciteBstWouldAddEndPuncttrue
\mciteSetBstMidEndSepPunct{\mcitedefaultmidpunct}
{\mcitedefaultendpunct}{\mcitedefaultseppunct}\relax
\EndOfBibitem
\bibitem[Witzler \latin{et~al.}(2018)Witzler, Alzagameem, Bergs, Khaldi-Hansen,
  Klein, Hielscher, Kamm, Kreyenschmidt, Tobiasch, and Schulze]{witzler2018}
Witzler,~M.; Alzagameem,~A.; Bergs,~M.; Khaldi-Hansen,~B.~E.; Klein,~S.~E.;
  Hielscher,~D.; Kamm,~B.; Kreyenschmidt,~J.; Tobiasch,~E.; Schulze,~M.
  Lignin-derived biomaterials for drug release and tissue engineering.
  \emph{Molecules} \textbf{2018}, \emph{23}, 1885\relax
\mciteBstWouldAddEndPuncttrue
\mciteSetBstMidEndSepPunct{\mcitedefaultmidpunct}
{\mcitedefaultendpunct}{\mcitedefaultseppunct}\relax
\EndOfBibitem
\bibitem[Yamada and Sixt(2019)Yamada, and Sixt]{yamada2019}
Yamada,~K.~M.; Sixt,~M. Mechanisms of 3D cell migration. \emph{Nat. Rev. Mol.
  Cell Biol.} \textbf{2019}, \emph{20}, 738--752\relax
\mciteBstWouldAddEndPuncttrue
\mciteSetBstMidEndSepPunct{\mcitedefaultmidpunct}
{\mcitedefaultendpunct}{\mcitedefaultseppunct}\relax
\EndOfBibitem
\bibitem[Markstedt \latin{et~al.}(2015)Markstedt, Mantas, Tournier,
  {Mart{\'{i}}nez {\'{A}}vila}, H{\"{a}}gg, and Gatenholm]{markstedt2015}
Markstedt,~K.; Mantas,~A.; Tournier,~I.; {Mart{\'{i}}nez {\'{A}}vila},~H.;
  H{\"{a}}gg,~D.; Gatenholm,~P. 3D bioprinting human chondrocytes with
  nanocellulose--alginate bioink for cartilage tissue engineering applications.
  \emph{Biomacromolecules} \textbf{2015}, \emph{16}, 1489--1496\relax
\mciteBstWouldAddEndPuncttrue
\mciteSetBstMidEndSepPunct{\mcitedefaultmidpunct}
{\mcitedefaultendpunct}{\mcitedefaultseppunct}\relax
\EndOfBibitem
\bibitem[Curvello \latin{et~al.}(2019)Curvello, Raghuwanshi, and
  Garnier]{curvello2019}
Curvello,~R.; Raghuwanshi,~V.~S.; Garnier,~G. Engineering nanocellulose
  hydrogels for biomedical applications. \emph{Adv. Colloid Interface Sci.}
  \textbf{2019}, \emph{267}, 47--61\relax
\mciteBstWouldAddEndPuncttrue
\mciteSetBstMidEndSepPunct{\mcitedefaultmidpunct}
{\mcitedefaultendpunct}{\mcitedefaultseppunct}\relax
\EndOfBibitem
\bibitem[Ahmed \latin{et~al.}(2020)Ahmed, Gultekinoglu, and
  Edirisinghe]{Ahmed2020}
Ahmed,~J.; Gultekinoglu,~M.; Edirisinghe,~M. {Bacterial cellulose micro-nano
  fibres for wound healing applications}. \emph{Biotechnol. Adv.}
  \textbf{2020}, \emph{41}, 107549\relax
\mciteBstWouldAddEndPuncttrue
\mciteSetBstMidEndSepPunct{\mcitedefaultmidpunct}
{\mcitedefaultendpunct}{\mcitedefaultseppunct}\relax
\EndOfBibitem
\bibitem[Gilbert \latin{et~al.}(2021)Gilbert, Tang, Ott, Dorr, Shaw, Sun, Lu,
  and Ellis]{gilbert2021living}
Gilbert,~C.; Tang,~T.-C.; Ott,~W.; Dorr,~B.~A.; Shaw,~W.~M.; Sun,~G.~L.;
  Lu,~T.~K.; Ellis,~T. Living materials with programmable functionalities grown
  from engineered microbial co-cultures. \emph{Nat. Mater.} \textbf{2021},
  \emph{20}, 691--700\relax
\mciteBstWouldAddEndPuncttrue
\mciteSetBstMidEndSepPunct{\mcitedefaultmidpunct}
{\mcitedefaultendpunct}{\mcitedefaultseppunct}\relax
\EndOfBibitem
\bibitem[Chen and Hu(2018)Chen, and Hu]{Chen2018a}
Chen,~C.; Hu,~L. {Nanocellulose toward Advanced Energy Storage Devices:
  Structure and Electrochemistry}. \emph{Acc. Chem. Res.} \textbf{2018},
  \emph{51}, 3154--3165\relax
\mciteBstWouldAddEndPuncttrue
\mciteSetBstMidEndSepPunct{\mcitedefaultmidpunct}
{\mcitedefaultendpunct}{\mcitedefaultseppunct}\relax
\EndOfBibitem
\bibitem[Kim \latin{et~al.}(2019)Kim, Lee, Lee, Chen, and Lee]{Kim2019}
Kim,~J.~H.; Lee,~D.; Lee,~Y.~H.; Chen,~W.; Lee,~S.~Y. {Nanocellulose for Energy
  Storage Systems: Beyond the Limits of Synthetic Materials}. \emph{Adv.
  Mater.} \textbf{2019}, \emph{31}, 1--16\relax
\mciteBstWouldAddEndPuncttrue
\mciteSetBstMidEndSepPunct{\mcitedefaultmidpunct}
{\mcitedefaultendpunct}{\mcitedefaultseppunct}\relax
\EndOfBibitem
\bibitem[Schaffner \latin{et~al.}(2017)Schaffner, R{\"u}hs, Coulter, Kilcher,
  and Studart]{schaffner2017}
Schaffner,~M.; R{\"u}hs,~P.~A.; Coulter,~F.; Kilcher,~S.; Studart,~A.~R. 3D
  printing of bacteria into functional complex materials. \emph{Sci. Adv.}
  \textbf{2017}, \emph{3}, eaao6804\relax
\mciteBstWouldAddEndPuncttrue
\mciteSetBstMidEndSepPunct{\mcitedefaultmidpunct}
{\mcitedefaultendpunct}{\mcitedefaultseppunct}\relax
\EndOfBibitem
\bibitem[Hausmann \latin{et~al.}(2018)Hausmann, Ruhs, Siqueira, Lauger,
  Libanori, Zimmermann, and Studart]{hausmann2018}
Hausmann,~M.~K.; Ruhs,~P.~A.; Siqueira,~G.; Lauger,~J.; Libanori,~R.;
  Zimmermann,~T.; Studart,~A.~R. Dynamics of cellulose nanocrystal alignment
  during 3D printing. \emph{ACS Nano} \textbf{2018}, \emph{12},
  6926--6937\relax
\mciteBstWouldAddEndPuncttrue
\mciteSetBstMidEndSepPunct{\mcitedefaultmidpunct}
{\mcitedefaultendpunct}{\mcitedefaultseppunct}\relax
\EndOfBibitem
\bibitem[Joshi \latin{et~al.}(2018)Joshi, Cook, and Mannoor]{joshi2018}
Joshi,~S.; Cook,~E.; Mannoor,~M.~S. Bacterial Nanobionics via 3D printing.
  \emph{Nano Lett.} \textbf{2018}, \emph{18}, 7448--7456\relax
\mciteBstWouldAddEndPuncttrue
\mciteSetBstMidEndSepPunct{\mcitedefaultmidpunct}
{\mcitedefaultendpunct}{\mcitedefaultseppunct}\relax
\EndOfBibitem
\bibitem[Quijano \latin{et~al.}(2021)Quijano, Speight, and
  Payne]{quijano2021future}
Quijano,~L.; Speight,~R.; Payne,~A. Future fashion, biotechnology and the
  living world: Microbial cell factories and forming new ‘oddkins’.
  \emph{Continuum} \textbf{2021}, \emph{35}, 897--913\relax
\mciteBstWouldAddEndPuncttrue
\mciteSetBstMidEndSepPunct{\mcitedefaultmidpunct}
{\mcitedefaultendpunct}{\mcitedefaultseppunct}\relax
\EndOfBibitem
\bibitem[da~Silva \latin{et~al.}(2021)da~Silva, de~Medeiros, de~Amorim,
  do~Nascimento, Converti, Costa, and Sarubbo]{da2021bacterial}
da~Silva,~C. J.~G.; de~Medeiros,~A.~D.; de~Amorim,~J. D.~P.;
  do~Nascimento,~H.~A.; Converti,~A.; Costa,~A. F.~S.; Sarubbo,~L.~A. Bacterial
  cellulose biotextiles for the future of sustainable fashion: a review.
  \emph{Environ. Chem. Lett.} \textbf{2021}, \emph{19}, 2967--2980\relax
\mciteBstWouldAddEndPuncttrue
\mciteSetBstMidEndSepPunct{\mcitedefaultmidpunct}
{\mcitedefaultendpunct}{\mcitedefaultseppunct}\relax
\EndOfBibitem
\bibitem[Rybchyn \latin{et~al.}(2021)Rybchyn, Biazik, Charlesworth, and
  le~Coutre]{rybchyn2021nanocellulose}
Rybchyn,~M.~S.; Biazik,~J.~M.; Charlesworth,~J.; le~Coutre,~J. Nanocellulose
  from Nata de Coco as a Bioscaffold for Cell-Based Meat. \emph{ACS Omega}
  \textbf{2021}, \emph{6}, 33923--33931\relax
\mciteBstWouldAddEndPuncttrue
\mciteSetBstMidEndSepPunct{\mcitedefaultmidpunct}
{\mcitedefaultendpunct}{\mcitedefaultseppunct}\relax
\EndOfBibitem
\bibitem[Witten and Ribbeck(2017)Witten, and Ribbeck]{Witten2017}
Witten,~J.; Ribbeck,~K. {The particle in the spider's web: transport through
  biological hydrogels}. \emph{Nanoscale} \textbf{2017}, \emph{9},
  8080--8095\relax
\mciteBstWouldAddEndPuncttrue
\mciteSetBstMidEndSepPunct{\mcitedefaultmidpunct}
{\mcitedefaultendpunct}{\mcitedefaultseppunct}\relax
\EndOfBibitem
\bibitem[Stylianopoulos \latin{et~al.}(2008)Stylianopoulos, Yeckel, Derby, Luo,
  Shephard, Sander, and Barocas]{stylianopoulos2008permeability}
Stylianopoulos,~T.; Yeckel,~A.; Derby,~J.~J.; Luo,~X.-J.; Shephard,~M.~S.;
  Sander,~E.~A.; Barocas,~V.~H. Permeability calculations in three-dimensional
  isotropic and oriented fiber networks. \emph{Phys. Fluids} \textbf{2008},
  \emph{20}, 123601\relax
\mciteBstWouldAddEndPuncttrue
\mciteSetBstMidEndSepPunct{\mcitedefaultmidpunct}
{\mcitedefaultendpunct}{\mcitedefaultseppunct}\relax
\EndOfBibitem
\bibitem[Stylianopoulos \latin{et~al.}(2010)Stylianopoulos, Diop-Frimpong,
  Munn, and Jain]{Stylianopoulos2010}
Stylianopoulos,~T.; Diop-Frimpong,~B.; Munn,~L.~L.; Jain,~R.~K. {Diffusion
  anisotropy in collagen gels and tumors: The effect of fiber network
  orientation}. \emph{Biophys. J.} \textbf{2010}, \emph{99}, 3119--3128\relax
\mciteBstWouldAddEndPuncttrue
\mciteSetBstMidEndSepPunct{\mcitedefaultmidpunct}
{\mcitedefaultendpunct}{\mcitedefaultseppunct}\relax
\EndOfBibitem
\bibitem[Sykes \latin{et~al.}(2016)Sykes, Dai, Sarsons, Chen, Rocheleau, Hwang,
  Zheng, Cramb, Rinker, and Chan]{sykes2016tailoring}
Sykes,~E.~A.; Dai,~Q.; Sarsons,~C.~D.; Chen,~J.; Rocheleau,~J.~V.;
  Hwang,~D.~M.; Zheng,~G.; Cramb,~D.~T.; Rinker,~K.~D.; Chan,~W.~C. Tailoring
  nanoparticle designs to target cancer based on tumor pathophysiology.
  \emph{Proc. Natl. Acad. Sci. U. S. A.} \textbf{2016}, \emph{113},
  E1142--E1151\relax
\mciteBstWouldAddEndPuncttrue
\mciteSetBstMidEndSepPunct{\mcitedefaultmidpunct}
{\mcitedefaultendpunct}{\mcitedefaultseppunct}\relax
\EndOfBibitem
\bibitem[Chauhan \latin{et~al.}(2011)Chauhan, Stylianopoulos, Boucher, and
  Jain]{Chauhan2011}
Chauhan,~V.~P.; Stylianopoulos,~T.; Boucher,~Y.; Jain,~R.~K. {Delivery of
  molecular and nanoscale medicine to tumors: transport barriers and
  strategies}. \emph{Annu. Rev. Chem. Biomol. Eng.} \textbf{2011}, \emph{2},
  281--298\relax
\mciteBstWouldAddEndPuncttrue
\mciteSetBstMidEndSepPunct{\mcitedefaultmidpunct}
{\mcitedefaultendpunct}{\mcitedefaultseppunct}\relax
\EndOfBibitem
\bibitem[Sentjabrskaja \latin{et~al.}(2016)Sentjabrskaja, Zaccarelli, {De
  Michele}, Sciortino, Tartaglia, Voigtmann, Egelhaaf, and
  Laurati]{Sentjabrskaja2016}
Sentjabrskaja,~T.; Zaccarelli,~E.; {De Michele},~C.; Sciortino,~F.;
  Tartaglia,~P.; Voigtmann,~T.; Egelhaaf,~S.~U.; Laurati,~M. {Anomalous
  dynamics of intruders in a crowded environment of mobile obstacles}.
  \emph{Nat. Commun.} \textbf{2016}, \emph{7}, 1--8\relax
\mciteBstWouldAddEndPuncttrue
\mciteSetBstMidEndSepPunct{\mcitedefaultmidpunct}
{\mcitedefaultendpunct}{\mcitedefaultseppunct}\relax
\EndOfBibitem
\bibitem[Poling-Skutvik \latin{et~al.}(2019)Poling-Skutvik, Roberts, Slim,
  Narayanan, Krishnamoorti, Palmer, and Conrad]{Poling-Skutvik2019}
Poling-Skutvik,~R.; Roberts,~R.~C.; Slim,~A.~H.; Narayanan,~S.;
  Krishnamoorti,~R.; Palmer,~J.~C.; Conrad,~J.~C. {Structure dominates
  localization of tracers within aging nanoparticle glasses}. \emph{J. Phys.
  Chem. Lett.} \textbf{2019}, \emph{10}, 1784--1789\relax
\mciteBstWouldAddEndPuncttrue
\mciteSetBstMidEndSepPunct{\mcitedefaultmidpunct}
{\mcitedefaultendpunct}{\mcitedefaultseppunct}\relax
\EndOfBibitem
\bibitem[Roberts \latin{et~al.}(2019)Roberts, Poling-Skutvik, Conrad, and
  Palmer]{roberts2019tracer}
Roberts,~R.~C.; Poling-Skutvik,~R.; Conrad,~J.~C.; Palmer,~J.~C. Tracer
  transport in attractive and repulsive supercooled liquids and glasses.
  \emph{J. Chem. Phys.} \textbf{2019}, \emph{151}, 194501\relax
\mciteBstWouldAddEndPuncttrue
\mciteSetBstMidEndSepPunct{\mcitedefaultmidpunct}
{\mcitedefaultendpunct}{\mcitedefaultseppunct}\relax
\EndOfBibitem
\bibitem[Jain and Stylianopoulos(2010)Jain, and Stylianopoulos]{Jain2010}
Jain,~R.~K.; Stylianopoulos,~T. {Delivering nanomedicine to solid tumors}.
  \emph{Nat. Rev. Clin. Oncol.} \textbf{2010}, \emph{7}, 653--664\relax
\mciteBstWouldAddEndPuncttrue
\mciteSetBstMidEndSepPunct{\mcitedefaultmidpunct}
{\mcitedefaultendpunct}{\mcitedefaultseppunct}\relax
\EndOfBibitem
\bibitem[Yu \latin{et~al.}(2018)Yu, Xu, Tian, Su, Zheng, Yang, Wang, Wang, Zhu,
  Guo, \latin{et~al.} others]{yu2018rapid}
Yu,~M.; Xu,~L.; Tian,~F.; Su,~Q.; Zheng,~N.; Yang,~Y.; Wang,~J.; Wang,~A.;
  Zhu,~C.; Guo,~S., \latin{et~al.}  Rapid transport of deformation-tuned
  nanoparticles across biological hydrogels and cellular barriers. \emph{Nat.
  Commun.} \textbf{2018}, \emph{9}, 1--11\relax
\mciteBstWouldAddEndPuncttrue
\mciteSetBstMidEndSepPunct{\mcitedefaultmidpunct}
{\mcitedefaultendpunct}{\mcitedefaultseppunct}\relax
\EndOfBibitem
\bibitem[Yu \latin{et~al.}(2019)Yu, Song, Tian, Dai, Zhu, Ahmad, Guo, Zhu,
  Zhong, Yuan, Zhang, Yi, Shi, Gan, and Gao]{Yu2019}
Yu,~M.; Song,~W.; Tian,~F.; Dai,~Z.; Zhu,~Q.; Ahmad,~E.; Guo,~S.; Zhu,~C.;
  Zhong,~H.; Yuan,~Y.; Zhang,~T.; Yi,~X.; Shi,~X.; Gan,~Y.; Gao,~H.
  {Temperature- and rigidity-mediated rapid transport of lipid nanovesicles in
  hydrogels}. \emph{Proc. Natl. Acad. Sci. U. S. A.} \textbf{2019}, \emph{116},
  5362--5369\relax
\mciteBstWouldAddEndPuncttrue
\mciteSetBstMidEndSepPunct{\mcitedefaultmidpunct}
{\mcitedefaultendpunct}{\mcitedefaultseppunct}\relax
\EndOfBibitem
\bibitem[Bao \latin{et~al.}(2020)Bao, Liu, Li, Chai, Zhang, Jiao, Li, Yu, Ren,
  Shi, and Li]{Bao2020}
Bao,~C.; Liu,~B.; Li,~B.; Chai,~J.; Zhang,~L.; Jiao,~L.; Li,~D.; Yu,~Z.;
  Ren,~F.; Shi,~X.; Li,~Y. {Enhanced transport of shape and rigidity-tuned
  $\alpha$-Lactalbumin nanotubes across intestinal mucus and cellular
  barriers}. \emph{Nano Lett.} \textbf{2020}, \emph{20}, 1352--1361\relax
\mciteBstWouldAddEndPuncttrue
\mciteSetBstMidEndSepPunct{\mcitedefaultmidpunct}
{\mcitedefaultendpunct}{\mcitedefaultseppunct}\relax
\EndOfBibitem
\bibitem[Huck \latin{et~al.}(2019)Huck, Hartwig, Biehl, Schwarzkopf, Wagner,
  Loretz, Murgia, and Lehr]{Huck2019}
Huck,~B.~C.; Hartwig,~O.; Biehl,~A.; Schwarzkopf,~K.; Wagner,~C.; Loretz,~B.;
  Murgia,~X.; Lehr,~C.~M. {Macro- and imcrorheological properties of mucus
  surrogates in comparison to native intestinal and pulmonary mucus}.
  \emph{Biomacromolecules} \textbf{2019}, \emph{20}, 3504--3512\relax
\mciteBstWouldAddEndPuncttrue
\mciteSetBstMidEndSepPunct{\mcitedefaultmidpunct}
{\mcitedefaultendpunct}{\mcitedefaultseppunct}\relax
\EndOfBibitem
\bibitem[Birjiniuk \latin{et~al.}(2014)Birjiniuk, Billings, Nance, Hanes,
  Ribbeck, and Doyle]{birjiniuk2014single}
Birjiniuk,~A.; Billings,~N.; Nance,~E.; Hanes,~J.; Ribbeck,~K.; Doyle,~P.~S.
  Single particle tracking reveals spatial and dynamic organization of the
  Escherichia coli biofilm matrix. \emph{New J. Phys.} \textbf{2014},
  \emph{16}, 085014\relax
\mciteBstWouldAddEndPuncttrue
\mciteSetBstMidEndSepPunct{\mcitedefaultmidpunct}
{\mcitedefaultendpunct}{\mcitedefaultseppunct}\relax
\EndOfBibitem
\bibitem[Billings \latin{et~al.}(2015)Billings, Birjiniuk, Samad, Doyle, and
  Ribbeck]{billings2015material}
Billings,~N.; Birjiniuk,~A.; Samad,~T.~S.; Doyle,~P.~S.; Ribbeck,~K. Material
  properties of biofilms—a review of methods for understanding permeability
  and mechanics. \emph{Rep. Prog. Phys.} \textbf{2015}, \emph{78}, 036601\relax
\mciteBstWouldAddEndPuncttrue
\mciteSetBstMidEndSepPunct{\mcitedefaultmidpunct}
{\mcitedefaultendpunct}{\mcitedefaultseppunct}\relax
\EndOfBibitem
\bibitem[Mastorakos \latin{et~al.}(2015)Mastorakos, Silva, Chisholm, Song,
  Choi, Boyle, Morales, Hanes, and Suk]{Mastorakos2015}
Mastorakos,~P.; Silva,~A.~L.; Chisholm,~J.; Song,~E.; Choi,~W.~K.;
  Boyle,~M.~P.; Morales,~M.~M.; Hanes,~J.; Suk,~J.~S. {Highly compacted
  biodegradable DNA nanoparticles capable of overcoming the mucus barrier for
  inhaled lung gene therapy}. \emph{Proc. Natl. Acad. Sci. U. S. A.}
  \textbf{2015}, \emph{112}, 8720--8725\relax
\mciteBstWouldAddEndPuncttrue
\mciteSetBstMidEndSepPunct{\mcitedefaultmidpunct}
{\mcitedefaultendpunct}{\mcitedefaultseppunct}\relax
\EndOfBibitem
\bibitem[Marczynski \latin{et~al.}(2018)Marczynski, K{\"{a}}sdorf, Altaner,
  Wenzler, Gerland, and Lieleg]{Marczynski2018}
Marczynski,~M.; K{\"{a}}sdorf,~B.~T.; Altaner,~B.; Wenzler,~A.; Gerland,~U.;
  Lieleg,~O. {Transient binding promotes molecule penetration into mucin
  hydrogels by enhancing molecular partitioning}. \emph{Biomater. Sci.}
  \textbf{2018}, \emph{6}, 3373--3387\relax
\mciteBstWouldAddEndPuncttrue
\mciteSetBstMidEndSepPunct{\mcitedefaultmidpunct}
{\mcitedefaultendpunct}{\mcitedefaultseppunct}\relax
\EndOfBibitem
\bibitem[Dunsing \latin{et~al.}(2019)Dunsing, Irmscher, Barbirz, and
  Chiantia]{dunsing2019purely}
Dunsing,~V.; Irmscher,~T.; Barbirz,~S.; Chiantia,~S. Purely
  polysaccharide-based biofilm matrix provides size-selective diffusion
  barriers for nanoparticles and bacteriophages. \emph{Biomacromolecules}
  \textbf{2019}, \emph{20}, 3842--3854\relax
\mciteBstWouldAddEndPuncttrue
\mciteSetBstMidEndSepPunct{\mcitedefaultmidpunct}
{\mcitedefaultendpunct}{\mcitedefaultseppunct}\relax
\EndOfBibitem
\bibitem[Stewart \latin{et~al.}(2015)Stewart, Ganesan, Younger, and
  Solomon]{stewart2015artificial}
Stewart,~E.~J.; Ganesan,~M.; Younger,~J.~G.; Solomon,~M.~J. Artificial biofilms
  establish the role of matrix interactions in staphylococcal biofilm assembly
  and disassembly. \emph{Sci. Rep.} \textbf{2015}, \emph{5}, 1--14\relax
\mciteBstWouldAddEndPuncttrue
\mciteSetBstMidEndSepPunct{\mcitedefaultmidpunct}
{\mcitedefaultendpunct}{\mcitedefaultseppunct}\relax
\EndOfBibitem
\bibitem[Ganesan \latin{et~al.}(2016)Ganesan, Knier, Younger, and
  Solomon]{ganesan2016associative}
Ganesan,~M.; Knier,~S.; Younger,~J.~G.; Solomon,~M.~J. Associative and
  entanglement contributions to the solution rheology of a bacterial
  polysaccharide. \emph{Macromolecules} \textbf{2016}, \emph{49},
  8313--8321\relax
\mciteBstWouldAddEndPuncttrue
\mciteSetBstMidEndSepPunct{\mcitedefaultmidpunct}
{\mcitedefaultendpunct}{\mcitedefaultseppunct}\relax
\EndOfBibitem
\bibitem[Kundukad \latin{et~al.}(2017)Kundukad, Schussman, Yang, Seviour, Yang,
  Rice, Kjelleberg, and Doyle]{kundukad2017mechanistic}
Kundukad,~B.; Schussman,~M.; Yang,~K.; Seviour,~T.; Yang,~L.; Rice,~S.~A.;
  Kjelleberg,~S.; Doyle,~P.~S. Mechanistic action of weak acid drugs on
  biofilms. \emph{Sci. Rep.} \textbf{2017}, \emph{7}, 1--12\relax
\mciteBstWouldAddEndPuncttrue
\mciteSetBstMidEndSepPunct{\mcitedefaultmidpunct}
{\mcitedefaultendpunct}{\mcitedefaultseppunct}\relax
\EndOfBibitem
\bibitem[Boudarel \latin{et~al.}(2021)Boudarel, Mathias, Blaysat, and
  Gr{\'e}diac]{boudarel2021situ}
Boudarel,~H.; Mathias,~J.-D.; Blaysat,~B.; Gr{\'e}diac,~M. In situ tracking of
  microbeads for the detection of biofilm formation. \emph{Biotechnol. Bioeng.}
  \textbf{2021}, \emph{118}, 1244--1261\relax
\mciteBstWouldAddEndPuncttrue
\mciteSetBstMidEndSepPunct{\mcitedefaultmidpunct}
{\mcitedefaultendpunct}{\mcitedefaultseppunct}\relax
\EndOfBibitem
\bibitem[Chew \latin{et~al.}(2014)Chew, Kundukad, Seviour, Van~der Maarel,
  Yang, Rice, Doyle, and Kjelleberg]{chew2014dynamic}
Chew,~S.~C.; Kundukad,~B.; Seviour,~T.; Van~der Maarel,~J.~R.; Yang,~L.;
  Rice,~S.~A.; Doyle,~P.; Kjelleberg,~S. Dynamic remodeling of microbial
  biofilms by functionally distinct exopolysaccharides. \emph{MBio}
  \textbf{2014}, \emph{5}, e01536--14\relax
\mciteBstWouldAddEndPuncttrue
\mciteSetBstMidEndSepPunct{\mcitedefaultmidpunct}
{\mcitedefaultendpunct}{\mcitedefaultseppunct}\relax
\EndOfBibitem
\bibitem[Kundukad \latin{et~al.}(2016)Kundukad, Seviour, Liang, Rice,
  Kjelleberg, and Doyle]{kundukad2016mechanical}
Kundukad,~B.; Seviour,~T.; Liang,~Y.; Rice,~S.~A.; Kjelleberg,~S.; Doyle,~P.~S.
  Mechanical properties of the superficial biofilm layer determine the
  architecture of biofilms. \emph{Soft matter} \textbf{2016}, \emph{12},
  5718--5726\relax
\mciteBstWouldAddEndPuncttrue
\mciteSetBstMidEndSepPunct{\mcitedefaultmidpunct}
{\mcitedefaultendpunct}{\mcitedefaultseppunct}\relax
\EndOfBibitem
\bibitem[Crocker \latin{et~al.}(1996)Crocker, Crocker, and Grier]{Crocker1996}
Crocker,~J.; Crocker,~J.; Grier,~D. {Methods of digital video microscopy for
  colloidal studies}. \emph{J. Colloid Interface Sci.} \textbf{1996},
  \emph{179}, 298--310\relax
\mciteBstWouldAddEndPuncttrue
\mciteSetBstMidEndSepPunct{\mcitedefaultmidpunct}
{\mcitedefaultendpunct}{\mcitedefaultseppunct}\relax
\EndOfBibitem
\bibitem[Bodin \latin{et~al.}(2007)Bodin, Backdahl, Fink, Gustafsson, Risberg,
  and Gatenholm]{Bodin2007}
Bodin,~A.; Backdahl,~H.; Fink,~H.; Gustafsson,~L.; Risberg,~B.; Gatenholm,~P.
  {Influence of Cultivation Conditions on Mechanical and Morphological
  Properties of Bacterial Cellulose Tubes}. \emph{Biotechnol. Bioeng.}
  \textbf{2007}, \emph{97}, 425--434\relax
\mciteBstWouldAddEndPuncttrue
\mciteSetBstMidEndSepPunct{\mcitedefaultmidpunct}
{\mcitedefaultendpunct}{\mcitedefaultseppunct}\relax
\EndOfBibitem
\bibitem[Hu \latin{et~al.}(2010)Hu, Gao, Tajima, Sunagawa, Zhou, Kawano,
  Fujiwara, Yoda, Shimura, Satoh, Munekata, Tanaka, and Yao]{Hu2010}
Hu,~S.~Q.; Gao,~Y.~G.; Tajima,~K.; Sunagawa,~N.; Zhou,~Y.; Kawano,~S.;
  Fujiwara,~T.; Yoda,~T.; Shimura,~D.; Satoh,~Y.; Munekata,~M.; Tanaka,~I.;
  Yao,~M. {Structure of bacterial cellulose synthase subunit D octamer with
  four inner passageways}. \emph{Proc. Natl. Acad. Sci. U. S. A.}
  \textbf{2010}, \emph{107}, 17957--17961\relax
\mciteBstWouldAddEndPuncttrue
\mciteSetBstMidEndSepPunct{\mcitedefaultmidpunct}
{\mcitedefaultendpunct}{\mcitedefaultseppunct}\relax
\EndOfBibitem
\bibitem[Janpetch \latin{et~al.}(2016)Janpetch, Saito, and
  Rujiravanit]{Janpetch2016}
Janpetch,~N.; Saito,~N.; Rujiravanit,~R. {Fabrication of bacterial
  cellulose-ZnO composite via solution plasma process for antibacterial
  applications}. \emph{Carbohydr. Polym.} \textbf{2016}, \emph{148},
  335--344\relax
\mciteBstWouldAddEndPuncttrue
\mciteSetBstMidEndSepPunct{\mcitedefaultmidpunct}
{\mcitedefaultendpunct}{\mcitedefaultseppunct}\relax
\EndOfBibitem
\bibitem[Gromovykh \latin{et~al.}(2020)Gromovykh, Pigaleva, Gallyamov,
  Ivanenko, Ozerova, Kharitonova, Bahman, Feldman, Lutsenko, and
  Kiselyova]{gromovykh2020structural}
Gromovykh,~T.~I.; Pigaleva,~M.~A.; Gallyamov,~M.~O.; Ivanenko,~I.~P.;
  Ozerova,~K.~E.; Kharitonova,~E.~P.; Bahman,~M.; Feldman,~N.~B.;
  Lutsenko,~S.~V.; Kiselyova,~O.~I. Structural organization of bacterial
  cellulose: The origin of anisotropy and layered structures. \emph{Carbohydr.
  Polym.} \textbf{2020}, \emph{237}, 116140\relax
\mciteBstWouldAddEndPuncttrue
\mciteSetBstMidEndSepPunct{\mcitedefaultmidpunct}
{\mcitedefaultendpunct}{\mcitedefaultseppunct}\relax
\EndOfBibitem
\bibitem[Qin \latin{et~al.}(2020)Qin, Fei, Bridges, Mashruwala, Stone,
  Wingreen, and Bassler]{qin2020cell}
Qin,~B.; Fei,~C.; Bridges,~A.~A.; Mashruwala,~A.~A.; Stone,~H.~A.;
  Wingreen,~N.~S.; Bassler,~B.~L. Cell position fates and collective fountain
  flow in bacterial biofilms revealed by light-sheet microscopy. \emph{Science}
  \textbf{2020}, \emph{369}, 71--77\relax
\mciteBstWouldAddEndPuncttrue
\mciteSetBstMidEndSepPunct{\mcitedefaultmidpunct}
{\mcitedefaultendpunct}{\mcitedefaultseppunct}\relax
\EndOfBibitem
\bibitem[Fu \latin{et~al.}(2012)Fu, Zhang, Li, Wu, Zhuo, Huang, Qiu, Zhou, and
  Yang]{Fu2012}
Fu,~L.; Zhang,~Y.; Li,~C.; Wu,~Z.; Zhuo,~Q.; Huang,~X.; Qiu,~G.; Zhou,~P.;
  Yang,~G. {Skin tissue repair materials from bacterial cellulose by a
  multilayer fermentation method}. \emph{J. Mater. Chem.} \textbf{2012},
  \emph{22}, 12349--12357\relax
\mciteBstWouldAddEndPuncttrue
\mciteSetBstMidEndSepPunct{\mcitedefaultmidpunct}
{\mcitedefaultendpunct}{\mcitedefaultseppunct}\relax
\EndOfBibitem
\bibitem[Huang \latin{et~al.}(2013)Huang, Chen, Nguyen, Tang, Zhang, and
  Yang]{Huang2013}
Huang,~L.; Chen,~X.; Nguyen,~T.~X.; Tang,~H.; Zhang,~L.; Yang,~G.
  {Nano-cellulose 3D-networks as controlled-release drug carriers}. \emph{J.
  Mater. Chem. B} \textbf{2013}, \emph{1}, 2976--2984\relax
\mciteBstWouldAddEndPuncttrue
\mciteSetBstMidEndSepPunct{\mcitedefaultmidpunct}
{\mcitedefaultendpunct}{\mcitedefaultseppunct}\relax
\EndOfBibitem
\bibitem[Zhou \latin{et~al.}(2007)Zhou, Sun, Hu, Li, and Yang]{Zhou2007}
Zhou,~L.~L.; Sun,~D.~P.; Hu,~L.~Y.; Li,~Y.~W.; Yang,~J.~Z. {Effect of addition
  of sodium alginate on bacterial cellulose production by Acetobacter xylinum}.
  \emph{J. Ind. Microbiol. Biotechnol.} \textbf{2007}, \emph{34},
  483--489\relax
\mciteBstWouldAddEndPuncttrue
\mciteSetBstMidEndSepPunct{\mcitedefaultmidpunct}
{\mcitedefaultendpunct}{\mcitedefaultseppunct}\relax
\EndOfBibitem
\bibitem[Cheng \latin{et~al.}(2009)Cheng, Catchmark, and Demirci]{Cheng2009}
Cheng,~K.~C.; Catchmark,~J.~M.; Demirci,~A. {Effect of different additives on
  bacterial cellulose production by Acetobacter xylinum and analysis of
  material property}. \emph{Cellulose} \textbf{2009}, \emph{16},
  1033--1045\relax
\mciteBstWouldAddEndPuncttrue
\mciteSetBstMidEndSepPunct{\mcitedefaultmidpunct}
{\mcitedefaultendpunct}{\mcitedefaultseppunct}\relax
\EndOfBibitem
\bibitem[Kwandou(2018)]{Kwandou2018}
Kwandou,~G. Characterization and control of biofilm and biofluid microstructure
  and rheology. Ph.D.\ thesis, University of New South Wales, Sydney,
  Australia, 2018\relax
\mciteBstWouldAddEndPuncttrue
\mciteSetBstMidEndSepPunct{\mcitedefaultmidpunct}
{\mcitedefaultendpunct}{\mcitedefaultseppunct}\relax
\EndOfBibitem
\bibitem[Sankaran \latin{et~al.}(2019)Sankaran, Tan, But, Cohen, Rice, and
  Wohland]{sankaran2019single}
Sankaran,~J.; Tan,~N.~J.; But,~K.~P.; Cohen,~Y.; Rice,~S.~A.; Wohland,~T.
  Single microcolony diffusion analysis in Pseudomonas aeruginosa biofilms.
  \emph{npj Biofilms and Microbiomes} \textbf{2019}, \emph{5}, 1--10\relax
\mciteBstWouldAddEndPuncttrue
\mciteSetBstMidEndSepPunct{\mcitedefaultmidpunct}
{\mcitedefaultendpunct}{\mcitedefaultseppunct}\relax
\EndOfBibitem
\bibitem[Remacha \latin{et~al.}(2020)Remacha, Friedrich, Vermot, and
  Fahrbach]{remacha2020define}
Remacha,~E.; Friedrich,~L.; Vermot,~J.; Fahrbach,~F.~O. How to define and
  optimize axial resolution in light-sheet microscopy: a simulation-based
  approach. \emph{Biomed. Opt. Express} \textbf{2020}, \emph{11}, 8--26\relax
\mciteBstWouldAddEndPuncttrue
\mciteSetBstMidEndSepPunct{\mcitedefaultmidpunct}
{\mcitedefaultendpunct}{\mcitedefaultseppunct}\relax
\EndOfBibitem
\bibitem[Cai \latin{et~al.}(2015)Cai, Panyukov, and Rubinstein]{cai2015hopping}
Cai,~L.-H.; Panyukov,~S.; Rubinstein,~M. Hopping diffusion of nanoparticles in
  polymer matrices. \emph{Macromolecules} \textbf{2015}, \emph{48},
  847--862\relax
\mciteBstWouldAddEndPuncttrue
\mciteSetBstMidEndSepPunct{\mcitedefaultmidpunct}
{\mcitedefaultendpunct}{\mcitedefaultseppunct}\relax
\EndOfBibitem
\bibitem[Valentine \latin{et~al.}(2001)Valentine, Kaplan, Thota, Crocker,
  Gisler, Prud’homme, Beck, and Weitz]{valentine2001investigating}
Valentine,~M.~T.; Kaplan,~P.~D.; Thota,~D.; Crocker,~J.~C.; Gisler,~T.;
  Prud’homme,~R.~K.; Beck,~M.; Weitz,~D.~A. Investigating the
  microenvironments of inhomogeneous soft materials with multiple particle
  tracking. \emph{Phys. Rev. E} \textbf{2001}, \emph{64}, 061506\relax
\mciteBstWouldAddEndPuncttrue
\mciteSetBstMidEndSepPunct{\mcitedefaultmidpunct}
{\mcitedefaultendpunct}{\mcitedefaultseppunct}\relax
\EndOfBibitem
\bibitem[Chaudhuri \latin{et~al.}(2007)Chaudhuri, Berthier, and
  Kob]{Chaudhuri2007}
Chaudhuri,~P.; Berthier,~L.; Kob,~W. {Universal nature of particle
  displacements close to glass and jamming transitions}. \emph{Phys. Rev.
  Lett.} \textbf{2007}, \emph{99}, 060604\relax
\mciteBstWouldAddEndPuncttrue
\mciteSetBstMidEndSepPunct{\mcitedefaultmidpunct}
{\mcitedefaultendpunct}{\mcitedefaultseppunct}\relax
\EndOfBibitem
\bibitem[Gao and Kilfoil(2007)Gao, and Kilfoil]{Gao2007}
Gao,~Y.; Kilfoil,~M.~L. {Direct imaging of dynamical heterogeneities near the
  colloid-gel transition}. \emph{Phys. Rev. Lett.} \textbf{2007}, \emph{99},
  1--4\relax
\mciteBstWouldAddEndPuncttrue
\mciteSetBstMidEndSepPunct{\mcitedefaultmidpunct}
{\mcitedefaultendpunct}{\mcitedefaultseppunct}\relax
\EndOfBibitem
\bibitem[Wang \latin{et~al.}(2012)Wang, Kuo, Bae, and Granick]{Wang2012}
Wang,~B.; Kuo,~J.; Bae,~S.~C.; Granick,~S. {When Brownian diffusion is not
  Gaussian.} \emph{Nat. Mater.} \textbf{2012}, \emph{11}, 481--5\relax
\mciteBstWouldAddEndPuncttrue
\mciteSetBstMidEndSepPunct{\mcitedefaultmidpunct}
{\mcitedefaultendpunct}{\mcitedefaultseppunct}\relax
\EndOfBibitem
\bibitem[Stuhrmann \latin{et~al.}(2012)Stuhrmann, {Soares E Silva}, Depken,
  MacKintosh, and Koenderink]{Stuhrmann2012}
Stuhrmann,~B.; {Soares E Silva},~M.; Depken,~M.; MacKintosh,~F.~C.;
  Koenderink,~G.~H. {Nonequilibrium fluctuations of a remodeling in vitro
  cytoskeleton}. \emph{Phys. Rev. E} \textbf{2012}, \emph{86}, 1--5\relax
\mciteBstWouldAddEndPuncttrue
\mciteSetBstMidEndSepPunct{\mcitedefaultmidpunct}
{\mcitedefaultendpunct}{\mcitedefaultseppunct}\relax
\EndOfBibitem
\bibitem[Lampo \latin{et~al.}(2017)Lampo, Stylianidou, Backlund, Wiggins, and
  Spakowitz]{Lampo2017}
Lampo,~T.~J.; Stylianidou,~S.; Backlund,~M.~P.; Wiggins,~P.~A.;
  Spakowitz,~A.~J. {Cytoplasmic RNA-protein particles exhibit non-Gaussian
  subdiffusive behavior}. \emph{Biophys. J.} \textbf{2017}, \emph{112},
  532--542\relax
\mciteBstWouldAddEndPuncttrue
\mciteSetBstMidEndSepPunct{\mcitedefaultmidpunct}
{\mcitedefaultendpunct}{\mcitedefaultseppunct}\relax
\EndOfBibitem
\bibitem[Kegel and van Blaaderen(2000)Kegel, and van
  Blaaderen]{kegel2000direct}
Kegel,~W.~K.; van Blaaderen,~A. Direct observation of dynamical heterogeneities
  in colloidal hard-sphere suspensions. \emph{Science} \textbf{2000},
  \emph{287}, 290--293\relax
\mciteBstWouldAddEndPuncttrue
\mciteSetBstMidEndSepPunct{\mcitedefaultmidpunct}
{\mcitedefaultendpunct}{\mcitedefaultseppunct}\relax
\EndOfBibitem
\bibitem[Wang \latin{et~al.}(2009)Wang, Anthony, Bae, and
  Granick]{wang2009anomalous}
Wang,~B.; Anthony,~S.~M.; Bae,~S.~C.; Granick,~S. Anomalous yet brownian.
  \emph{Proc. Natl. Acad. Sci. U. S. A.} \textbf{2009}, \emph{106},
  15160--15164\relax
\mciteBstWouldAddEndPuncttrue
\mciteSetBstMidEndSepPunct{\mcitedefaultmidpunct}
{\mcitedefaultendpunct}{\mcitedefaultseppunct}\relax
\EndOfBibitem
\bibitem[Weeks \latin{et~al.}(2000)Weeks, Crocker, Levitt, Schofield, and
  Weitz]{weeks2000three}
Weeks,~E.~R.; Crocker,~J.~C.; Levitt,~A.~C.; Schofield,~A.; Weitz,~D.~A.
  Three-dimensional direct imaging of structural relaxation near the colloidal
  glass transition. \emph{Science} \textbf{2000}, \emph{287}, 627--631\relax
\mciteBstWouldAddEndPuncttrue
\mciteSetBstMidEndSepPunct{\mcitedefaultmidpunct}
{\mcitedefaultendpunct}{\mcitedefaultseppunct}\relax
\EndOfBibitem
\bibitem[Ghazaryan \latin{et~al.}(2012)Ghazaryan, Tsai, Hayrapetyan, Chen,
  Chen, Jeong, Kim, Chen, and Dong]{ghazaryan2012analysis}
Ghazaryan,~A.; Tsai,~H.~F.; Hayrapetyan,~G.; Chen,~W.-L.; Chen,~Y.-F.;
  Jeong,~M.-Y.; Kim,~C.-S.; Chen,~S.-J.; Dong,~C.-Y. Analysis of collagen fiber
  domain organization by Fourier second harmonic generation microscopy.
  \emph{J. Biomed. Opt.} \textbf{2012}, \emph{18}, 031105\relax
\mciteBstWouldAddEndPuncttrue
\mciteSetBstMidEndSepPunct{\mcitedefaultmidpunct}
{\mcitedefaultendpunct}{\mcitedefaultseppunct}\relax
\EndOfBibitem
\bibitem[Soumpasis(1983)]{soumpasis1983theoretical}
Soumpasis,~D. Theoretical analysis of fluorescence photobleaching recovery
  experiments. \emph{Biophys. J.} \textbf{1983}, \emph{41}, 95--97\relax
\mciteBstWouldAddEndPuncttrue
\mciteSetBstMidEndSepPunct{\mcitedefaultmidpunct}
{\mcitedefaultendpunct}{\mcitedefaultseppunct}\relax
\EndOfBibitem
\bibitem[Carnell \latin{et~al.}(2015)Carnell, Macmillan, and
  Whan]{carnell2015fluorescence}
Carnell,~M.; Macmillan,~A.; Whan,~R. \emph{Methods in membrane lipids};
  Springer, 2015; pp 255--271\relax
\mciteBstWouldAddEndPuncttrue
\mciteSetBstMidEndSepPunct{\mcitedefaultmidpunct}
{\mcitedefaultendpunct}{\mcitedefaultseppunct}\relax
\EndOfBibitem
\bibitem[Hsu \latin{et~al.}(2018)Hsu, Wei, Guo, Phan, Zhang, and Chen]{Hsu2018}
Hsu,~M.~N.; Wei,~S.~C.; Guo,~S.; Phan,~D.~T.; Zhang,~Y.; Chen,~C.~H. {Smart
  hydrogel microfluidics for single-cell multiplexed secretomic analysis with
  high sensitivity}. \emph{Small} \textbf{2018}, \emph{14}, 1--11\relax
\mciteBstWouldAddEndPuncttrue
\mciteSetBstMidEndSepPunct{\mcitedefaultmidpunct}
{\mcitedefaultendpunct}{\mcitedefaultseppunct}\relax
\EndOfBibitem
\bibitem[Pedron \latin{et~al.}(2015)Pedron, Becka, and Harley]{Pedron2015}
Pedron,~S.; Becka,~E.; Harley,~B.~A. {Spatially gradated hydrogel platform as a
  3D engineered tumor microenvironment}. \emph{Adv. Mater.} \textbf{2015},
  \emph{27}, 1567--1572\relax
\mciteBstWouldAddEndPuncttrue
\mciteSetBstMidEndSepPunct{\mcitedefaultmidpunct}
{\mcitedefaultendpunct}{\mcitedefaultseppunct}\relax
\EndOfBibitem
\bibitem[Deforest and Tirrell(2015)Deforest, and Tirrell]{Deforest2015}
Deforest,~C.~A.; Tirrell,~D.~A. {A photoreversible protein-patterning approach
  for guiding stem cell fate in three-dimensional gels}. \emph{Nat. Mater.}
  \textbf{2015}, \emph{14}, 523--531\relax
\mciteBstWouldAddEndPuncttrue
\mciteSetBstMidEndSepPunct{\mcitedefaultmidpunct}
{\mcitedefaultendpunct}{\mcitedefaultseppunct}\relax
\EndOfBibitem
\bibitem[Loebel \latin{et~al.}(2020)Loebel, Kwon, Wang, Han, Mauck, and
  Burdick]{Loebel2020}
Loebel,~C.; Kwon,~M.~Y.; Wang,~C.; Han,~L.; Mauck,~R.~L.; Burdick,~J.~A.
  {Metabolic labeling to probe the spatiotemporal accumulation of matrix at the
  chondrocyte–hydrogel interface}. \emph{Adv. Funct. Mater.} \textbf{2020},
  \emph{30}, 1--10\relax
\mciteBstWouldAddEndPuncttrue
\mciteSetBstMidEndSepPunct{\mcitedefaultmidpunct}
{\mcitedefaultendpunct}{\mcitedefaultseppunct}\relax
\EndOfBibitem
\end{mcitethebibliography}

\end{document}